\documentclass[12pt]{article}

\usepackage{latexsym,amsbsy,amssymb,amsthm,amsmath,multirow}
\usepackage{epsfig}
\usepackage{natbib,graphicx,setspace,lscape,longtable}
\usepackage{color}
\usepackage{array}
\RequirePackage[mathlines, displaymath]{lineno}
\newcommand{\csection}[1]
    {\begin{center}
        \stepcounter{section}
        {\bf\large\arabic{section}. #1}
    \end{center}
    \vspace{-0.15 cm}
}

\newcommand{\csubsection}[1]{
\vspace{0.25 cm}
\noindent\stepcounter{subsection}
{\bf\arabic{section}.\arabic{subsection}. #1}
\vspace{0.25 cm}
}

\newtheorem{theorem}{Theorem}[section]
\newtheorem{proposition}{{\bf Proposition}}[section]
\newtheorem{lemma}{{\bf Lemma}}

\def\Y{Y}
\def\X{{\bf X}}
\def\Z{{\bf Z}}
\def\M{{\bf M}}
\def\W{{\bf W}}
\def\U{{\bf U}}
\def\V{{\bf V}}

\def\x{x}
\def\I{{\bf I}}
\def\0{{\bf 0}}
\def\R{\mathbb{R}}

\def\cala{\mathcal{A}}
\def\cali{\mathcal{I}}
\def\cals{\mathcal{S}}
\def\calf{\mathcal{F}}
\def\calu{\mbox{\boldmath$\mathcal{U}$}}
\newcommand{\calO}{{\mathcal O}}
\newcommand{\calH}{{\mathcal H}}

\def\xa{\X_{\mathcal{A}}}
\def\xf{\X_{\mathcal{F}}}
\def\xac{\X_{{\mathcal{A}}^c}}

\def\txf{\tilde\X_{\mathcal{F}}}

\def\ba{\mbox{\boldmath$\alpha$}}
\def\bfbeta{\mbox{\boldmath$\beta$}}
\def\bfeta{\mbox{\boldmath$\eta$}}
\def\bfdelta{\mbox{\boldmath$\delta$}}
\def\bfphi{\mbox{\boldmath$\phi$}}
\def\bfiota{\mbox{\boldmath$\iota$}}
\def\bfnu{\mbox{\boldmath$\nu$}}
\def\bftheta{\mbox{\boldmath$\vartheta$}}

\def\sig{{\bf \Sigma}}

\def\halfinvsig{\sig^{-1/2}}

\newcommand{\indep}{\;\, \rule[0em]{.03em}{.67em} \hspace{-.25em}
\rule[0em]{.65em}{.03em} \hspace{-.25em}
\rule[0em]{.03em}{.67em}\;\,}

\newcommand{\spc}{{\mathcal S}}
\newcommand{\spn}{\mathrm{Span}}

\newcommand{\var}{\mathrm{Var}}
\newcommand{\cov}{\mathrm{Cov}}
\def\trace{\mathrm{tr}}
\def\tr{\mathrm{tr}}
\def\argmin{\mathrm{argmin}}

\def\sir{\mathrm{\tiny SIR}}
\def\save{\mathrm{\tiny SAVE}}
\def\dr{\mathrm{\tiny DR}}

\def\eop{\hfill $\Box$ \\}
\def\A{{\bf A}}
\def\C{{\bf C}}

\def\D{{\bf D}}
\def\B{{\bf B}}
\def\bfpsi{\mbox{\boldmath$\psi$}}
\def\bfell{{\bf L}}
\def\sigf{\sig_{\mathcal{F}}}

\numberwithin{equation}{section}

\begin{document}
%\linenumbers
%
\begin{center}
{\Large\bf Trace Pursuit: A General Framework for Model-Free Variable Selection}\\
\bigskip
%{\sc }
 \footnotetext[3]{\textit{Address for correspondence:}
Li-Xing Zhu, Department of Mathematics,
The Hong Kong Baptist University,
Kowloon Tong, Hong Kong.
Email:  lzhu@hkbu.edu.hk.}

%{\sc Zhou Yu$^*$, Yuexiao Dong$^\dag$, and Mian Huang$^\ddag$}
Zhou Yu$^*$, Yuexiao Dong$^\dag$, and Li-Xing Zhu$^\ddag$

$^*$East China Normal University,\\
$^\dag$Temple University, and
\\$^\ddag$The Hong Kong Baptist University

\date{Feb 18, 2014}
\end{center}

\newpage

\begin{center}
{\Large\bf Trace Pursuit: A General Framework for Model-Free Variable Selection}\\
\bigskip
%%{\sc }
% \footnotetext[2]{\textit{Address for correspondence:}
%Li-Xing Zhu, Department of Mathematics,
%The Hong Kong Baptist University,
%Kowloon Tong, Hong Kong.
%Email:  lzhu@hkbu.edu.hk.}
%
%%{\sc Zhou Yu$^*$, Yuexiao Dong$^\dag$, and Mian Huang$^\ddag$}
%Zhou Yu$^*$, Yuexiao Dong$^\dag$, and Li-Xing Zhu$^\ddag$
%
%$^*$East China Normal University,\\
%$^\dag$Temple University, and
%\\$^\ddag$The Hong Kong Baptist University

%\today
\end{center}

\begin{abstract}
We propose trace pursuit for model-free variable selection under the sufficient dimension reduction paradigm.
Two distinct algorithms are proposed: stepwise trace pursuit and forward trace pursuit.
%, both of which can be combined with many existing sufficient dimension reduction methods.
Stepwise trace pursuit achieves selection consistency with fixed $p$, and is readily applicable in the challenging setting with $p>n$.
Forward trace pursuit can serve as an initial screening step to speed up the computation in the case of ultrahigh dimensionality.
The screening consistency property of forward trace pursuit based on sliced inverse regression is established.  Finite sample performances of trace pursuit and other model-free variable selection methods are compared through numerical studies.

\end{abstract}
\newpage

\csection{Introduction}
\label{sec:intro}

Contemporary statistical analysis often encounters high dimensional
datasets that are routinely collected in a wide range of research
areas, where the predictor dimensionality may easily dominate the
relatively small sample size. To include the significant variables
and exclude the insignificant variables at the same time, the
paradigm of variable selection has seen much progress in recent
years. Many popular variable selection procedures are developed
under the linear model or the generalized linear model framework, such as nonnegative
garrotte (Breiman, 1995), LASSO (Tibshirani, 1996),
SCAD (Fan and Li, 2001),  adaptive LASSO (Zou, 2006),  group LASSO (Yuan and Lin, 2006),  Dantzig selector
(Cand\'es and Tao, 2007), and MCP (Zhang, 2010).

Let
$\X=(\x_1,\cdots,\x_p)^T$ be the predictor and $\Y$ be the scalar
response. Model-free
variable selection
   aims to find the index set $\cala$ such that
\begin{align}\label{mfvs}
\Y \indep \xac|\xa,
\end{align}
  where  $\indep$ stands for independence, $\cala^c$ is the complement of $\cala$ in the index set
$\cali=\{1,\cdots,p\}$,
$\xa=\{\x_i: i\in \cala\}$, and $\xac=\{\x_i: i\in \cala^c\}$. Condition (\ref{mfvs}) implies that $\xa$
contains all the active predictors in terms of predicting $\Y$.
Ideally, we want to find the smallest index set $\cala$ satisfying
(\ref{mfvs}), in which case no inactive predictors are included in
$\xa$. Model-free variable selection is closely related to sufficient
dimension reduction (Li, 1991; Cook,
1998), which aims to find subspace $\cals$ such that
\begin{align}\label{SDR}
\Y \indep \X  | P_\cals \X.
\end{align}
Here  $P_{(\cdot)}$ denotes the projection operator with respect to
the standard inner product.
% (\ref{SDR}) implies that the conditional distribution of $\Y|\X$ is the same as that of $\Y| P_\cals \X$. Compared with $\X$, the projected  predictor $P_\cals \X$ has potentially much smaller dimensionality while retains all the regression information for response $\Y$.
Under mild conditions (Yin et al., 2008), the intersection of
all $\cals$ satisfying (\ref{SDR}) still satisfies (\ref{SDR}). We
call this intersection the central space and denote it by
$\cals_{\Y|\X}$. The dimension of $\cals_{\Y|\X}$ is called the
structural dimension and we denote it by $q$ with $q<p$. Some popular
sufficient dimension reduction methods in the literature include
sliced inverse regression (SIR) (Li, 1991), sliced average variance
estimation (SAVE) (Cook and Weisberg, 1991),
 %principal Hessian Direction (PHD) (Li, 1992; Cook, 1998b),
 and directional regression  (Li and Wang, 2007).

There are two distinct approaches in the literature for model-free variable selection: the sparse
sufficient dimension reduction approach and the hypothesis testing
approach.
By noting that many dimension reduction methods could be
reformulated as a least squares problem, Li (2007) proposed sparse
sufficient dimension reduction by combining sufficient dimension
reduction with penalized least squares. Other sparse dimension
reduction methods include shrinkage SIR (Ni et al., 2005),
constrained canonical correlation (Zhou and He, 2008), and
regularized SIR (Li and Yin, 2008). While traditional sufficient
dimension reduction finds linear combinations of all the original
variables, sparse sufficient dimension reduction achieves dimension
reduction and variable selection simultaneously.  The state of the
art method in this category is the coordinate-independent sparse
dimension reduction (CISE) (Chen et al., 2010), where a
subspace-oriented penalty is proposed such that the resulting
central space has the same sparsity structure regardless of the
chosen basis of $\cals_{\Y|\X}$. Although it enjoys the oracle property that it performs
asymptotically as well as if the true irrelevant predictors were
known, CISE is not applicable when $p$
is larger than the sample size $n$.
%the predictor dimensionality $p$
%is larger than the sample size $n$.

Model-free variable selection through sufficient dimension reduction
can also be implemented under the hypothesis testing framework.
Without loss of generality, we assume the active index set
$\cala=\{1,\cdots,K\}$ for  ease of demonstration.   Then (\ref{mfvs}) is
equivalent to $P_\calH \spc_{\Y|\X}=\calO_p$, where $\calH=\spn\left\{(\0_{(p-K)\times
K},\I_{p-K})^T\right\}$ is the subspace of $\R^p$ corresponding to
the coordinates of the inactive predictors, and $\calO_p$
denotes the origin in $\R^p$. To test $H_0: P_\calH
\spc_{\Y|\X}=\calO_p$ versus $H_a: P_\calH \spc_{\Y|\X}\neq\calO_p$,
Cook (2004) proposed the marginal coordinate hypothesis test based
on SIR,  and a similar test based on SAVE was developed in Shao et al. (2007). Backward elimination for variable selection
based on such tests is discussed in Li et al. (2005). However, these tests rely on an initial estimator
of the central space $\spc_{\Y|\X}$ via SIR or SAVE, which is not available when $p>n$.

To achieve model-free variable selection with $p>n$,
 Zhong el al. (2012) proposed correlation pursuit (COP). COP looks for a subset of variables in $\X$ to maximize an objective function, which measures
the correlation between the transformed response $\Y$ and the projections of $\X$. COP is based on SIR and inherits the limitations of SIR. Namely, COP may miss significant predictors linked to the response
through quadratic functions or interactions. More recently, Jiang and Liu (2013) proposed a likelihood ratio test based procedure named SIR with interaction detection (SIRI). SIRI includes a special case that is asymptotically equivalent to COP, and it extends COP by detecting significant predictors that appear in interactions.
Both COP and SIRI involve estimation of the structural dimension $q$ of $\spc_{\Y|\X}$, which
is known as order determination in the sufficient dimension reduction literature. Order determination
in the $p>n$ setting is a challenging issue, and the performances of COP and SIRI may deteriorate when the structural dimension $q$ can not be accurately estimated.

We propose trace pursuit as a novel approach for model-free variable selection in this paper.
Based on the newly designed method-specific (SIR, SAVE, or directional
regression) trace tests, we first extend the classical stepwise regression in linear models
and propose a  stepwise trace pursuit (STP) algorithm for model-free variable selection.
%STP starts from comparing $p$ traces of $1\times 1$ kernel matrices
%and adding the most significant predictor to the working index set
%$\calf$. In subsequent steps,
STP iterates between adding one
predictor from outside the working index set $\calf$ and deleting one
predictor from within $\calf$. Furthermore, we mimic the forward regression in the linear model setting and propose
the forward trace pursuit (FTP) algorithm. After finding a solution path by adding one predictor into the model at a time, a modified BIC
criterion provides a chosen model that is guaranteed to include all important predictors.
%Following the idea of Fan and Lv (2008), our
Finally, our two-stage hybrid trace pursuit (HTP) algorithm uses
FTP for initial variable screening, which is
followed by STP for the refined variable selection at the second
stage.  While SIR-based HTP
might miss some significant predictors involved only in interactions and
SAVE-based HTP may miss significant predictors that is linked to the
response through a linear function,
HTP based on directional regression   can successfully detect predictors in a wide range of models.
Compared with existing methods in the literature, the trace pursuit:  (1)  can be combined with different existing sufficient dimension reduction methods to  detect significant predictors linked through various unknown functions to the response; (2) does not rely on estimation of the structural dimension $q$; (3)
 is designed to deal with the challenging $p>n$ setting; (4) provides a unified framework for model-free variable screening through FTP and model-free variable selection through STP.
The selection consistency of the STP algorithms as well as the screening consistency property of the SIR-based FTP algorithm are established.

The paper is organized as follows.
%We review SIR, SAVE and DR for sufficient dimension reduction in Section 2.
We first propose the SIR-based trace test  and then extend it to SAVE-based and
directional regression-based trace tests in Section 2. The asymptotic distributions of
the proposed test statistics are discussed in Section 3. The STP algorithm and its selection consistency property   are developed in Section
4. FTP for screening and HTP for two-stage model-free variable selection are discussed in Section 5. Section
6 provides some numerical studies including a real data analysis.
Section 7 concludes the paper with some discussions.

\csection{Principle of the trace test}

\csubsection{Some preliminaries}

We briefly review three popular sufficient dimension reduction methods, SIR, SAVE and directional regression.
 Without loss of generality, assume $E(\X)=\0$ and $E(\Y)=0$.
% We denote the structural dimension of $\spc_{\Y|\X}$ as $d$, and
%the basis of $\spc_{\Y|\X}$ as .
Let $\var(\X)=\sig$ and $\Z=\sig^{-1/2}\X$ denotes the standardized
predictor. Suppose $\bfbeta\in\R^{p\times q}$ is the basis of
$\spc_{\Y|\X}$ and $\bfeta\in\R^{p\times q}$ is the basis of the
$\Z$-scaled central space $\spc_{\Y|\Z}$.
Let $\{J_1, \ldots ,J_H\}$ be a measurable partition of
$\Omega_{\Y}$, the sample space of $\Y$.
The kernel matrix of the classical SIR (Li, 1991) is defined as
$\M^\sir=\var \left\{ E(\Z | \Y\in J_h )\right\}$.
 Under the linear conditional mean (LCM) assumption that
\begin{align}
\label{eq:lcm} E(\X | \bfbeta^T \X) \mbox{ is a linear function of }
\bfbeta^T \X,
\end{align}
we have
%$E(\Z|\Y)\subseteq \spc_{\Y|\Z}$ and
$\spn(\M^\sir)\subseteq \spc_{\Y|\Z}$. Here $\spn(\M)$ denotes
 the column space of $\M$.
  Under the additional
 constant conditional variance (CCV) assumption that
\begin{align}
\label{eq:ccv} \var(\X | \bfbeta^T \X) \mbox{ is nonrandom},
\end{align}
 Cook and Weisberg (1991)
demonstrate that $\spn(\M^\save)$ $\subseteq \spc_{\Y|\Z}$, where  $\M^\save= E
\left\{ \I_p-\var(\Z |  \Y\in J_h )\right\}^2$ is the
 kernel matrix for SAVE.
  When $\X$ is
normal, both LCM and CCV assumptions are satisfied. For nonnormal predictor $\X$, please refer to
Cook and Nachtsheim (1994),  Cook and Li (2009), Li and Dong (2009), Dong
and Li (2010).

It is well-known that SIR and SAVE are complement to each other in both the regression and the classification settings. SIR works better when
the link function between the continuous response and the predictor is monotone, or when there is location shift between different categories of the discrete response. SAVE, on the other hand, is more effective with U-shaped link
function or detecting scale difference. Directional regression  is designed to combine the strength of SIR and SAVE.
For kernel matrix
\begin{align*}
&\M^\dr= 2 E\{E^2(\Z\Z^T|  \Y\in J_h) \}  +2 E^2\{E(\Z|\Y\in J_h)E^T(\Z|\Y\in J_h)\}\\
&\hspace{.2in}+2 E\{E^T(\Z|\Y\in J_h)E(\Z|\Y\in J_h)\}  E\{E(\Z|\Y\in J_h)E^T(\Z|\Y\in J_h)\}-2\I_p,
\end{align*}
Li and Wang (2007) prove that $\spn(\M^\dr)\subseteq \spc_{\Y|\Z}$ under assumptions  (\ref{eq:lcm}) and (\ref{eq:ccv}).

\csubsection{SIR-based trace test }
\label{sec:sir}

We state the principle of the SIR-based trace test in this section.
For working index set $\calf$ and index $j\in\calf^c$, we want to test
 \begin{align}\label{distribution test}
H_0: \Y\indep \x_j| \xf  \mbox{ v.s. }H_a: \Y\mbox{ is not
independent of } \x_j \mbox{ given } \xf.
\end{align}

 Denote $R_h=I(\Y\in J_h)$,
$p_h=E(R_h)$, and $\U_h=E(\X|\Y \in J_h)$. The
 kernel matrix for SIR can be rewritten as $\M^\sir=\halfinvsig\left(\sum_{h=1}^H p_h  \U_h \U_h^T\right)\halfinvsig$.
 For any index set $\calf$, denote $\xf=\{\x_i: i\in \calf\}$,
$\var(\xf)=\sig_{\calf}$, and $\U_{\calf,h}=E(\xf|\Y \in J_h)$.
 % Without loss of generality, let
%Some notations are needed before we introduce the trace test based
%on SIR. Suppose $\{J_1, \ldots ,J_H\}$ is a (measurable) partition of
%$\Omega_{\Y}$, the sample space of $\Y$.  Denote $R_h=I(\Y\in J_h)$,
%$p_h=E(R_h)$, and $\U_h=E(\X|\Y \in J_h)$.
%%Then $\delta(\Y) = \sum_{h=1}^H h I(\Y \in J_h)$ is the discretized version of $\Y$.
%With slight abuse of notation, the
% kernel matrix for the discretized SIR
%% $\M^\sir=   \var \left\{ E(\Z | \Y\in J_h )\right\}=\sig^{-1/2}\var \left\{ E(\Z | \Y\in J_h )\right\}\sig^{-1/2}$. . Then we
% can be written as $$\M^\sir=\halfinvsig\left(\sum_{h=1}^H p_h  \U_h
%\verb""\U_h^T\right)\halfinvsig.$$
%% \begin{align*}
%%\M^\sir=\halfinvsig\left(\sum_{h=1}^H p_h  U_h
%%U_h^T\right)\halfinvsig.
%%\end{align*}
%For a working index set $\calf$, denote $\xf=\{\x_i: i\in \calf\}$,
%$\var(\xf)=\sig_{\calf}$, and $\U_{\calf,h}=E(\xf|\Y \in J_h)$. We
%mimic $\M^\sir$  and
%Define $\M_{\calf}^\sir$ as
%\begin{align}
%\label{eq:kernel sir}
%\M_{\calf}^\sir=\sig_{\calf}^{-1/2}\left(\sum_{h=1}^H p_h
%\U_{\calf,h}\U_{\calf,h}^T\right)\sig_{\calf}^{-1/2}.
%\end{align}
We mimic $\M^\sir$ and define $\M_{\calf}^\sir$ as
\begin{align}
\label{eq:kernel sir}
\M_{\calf}^\sir=\sig_{\calf}^{-1/2}\left(\sum_{h=1}^H p_h
\U_{\calf,h}\U_{\calf,h}^T\right)\sig_{\calf}^{-1/2}.
\end{align}
Recall that $\cala$ denotes the active index set satisfying $\Y \indep \xac|\xa$, and
$\cali=\{1,\ldots,p\}$ denotes the full index set.
We have the following key observation.

\begin{proposition}
\label{prop:trace}
Suppose the LCM assumption  (\ref{eq:lcm}) holds true. Then for any index set
$\calf$ such that $\cala\subseteq \calf \subseteq \cali$, we have
$\trace(\M^\sir_\cala)=\trace(\M^\sir_\calf)=\trace(\M^\sir)$.
\end{proposition}

Proposition \ref{prop:trace} suggests that we use
$\trace(\M_{\calf}^\sir)$ to capture the strength of relationship
between $\Y$ and $\xf$.  Denote
 $\calf\cup j$ as the index set of $j$ together with all the indices in $\calf$.
 Given that $\xf$ is already in the model, we will see that the trace difference $\trace(\M^\sir_{\calf\cup j})-\trace(\M^\sir_\calf)$
 can be used to test the contribution of the additional variable $\x_j$ to $Y$.
 The  following subset LCM assumption is required before we state the main result,
 \begin{align}
\label{eq:subset lcm} E(\x_j | \xf) \mbox{ is a linear function of }
\xf \mbox{ for any }\calf\subset \cali \mbox { and } j\in\calf^c.
\end{align}
  Assumption (\ref{eq:subset lcm}) is parallel to the LCM assumption (\ref{eq:lcm}), and both are satisfied when $\X$ is elliptically contour
  distributed.
 The principle of the SIR-based trace test is stated in the next theorem.
 \begin{theorem}\label{prop:sir}
 Assume the
subset LCM assumption (\ref{eq:subset lcm})  holds true.
 Then for  $\calf\subset \cali$ and $j\in \calf^c$, we have
\begin{enumerate}
\item  $\trace(\M^\sir_{\calf\cup j})-\trace(\M^\sir_\calf)=\sum_{h=1}^H p_h
\gamma_{j|\mathcal{F},h}^2$, where
$\gamma_{j|\mathcal{F},h}=E(\gamma_{j|\mathcal{F}}|\Y \in J_h)$ with
$\x_{j|\mathcal{F}}=\x_j-E(\x_j|\xf)$,
$\sigma_{j|\mathcal{F}}^2=\var(\x_{j|\mathcal{F}})$, and
$\gamma_{j|\mathcal{F}}=\x_{j|\mathcal{F}}/\sigma_{j|\mathcal{F}}$.
%\mbox{ for any }\calf\subset \cali \mbox { and } j\in \calf^c$.
\item  $\trace(\M^\sir_{\calf\cup j})-\trace(\M^\sir_\calf)=0$ given that
%$\Y\indep \x_j| \xf$ and
$\cala\subseteq\calf$.
\end{enumerate}
\end{theorem}
\noindent
 Part 1 of Theorem \ref{prop:sir} provides the explicit formula to calculate the trace difference between $\M_{\calf\cup j}^\sir$ and $\M_{\calf}^\sir$. Part 2 of Theorem \ref{prop:sir} states that the trace difference is zero when the working index set $\calf$ contains the active set $\cala$.

% and Proposition \ref{prop:trace} have similar results, but they depend on different assumptions. Part 2 of Theorem \ref{prop:sir}
%, which is critical for the asymptotic analysis of the correponding sample version estimator.

%To test the contribution of the additional variable $\x_j$ to $Y$ given that $\xf$ is already in the model,
%Theorem \ref{prop:sir}  suggests that we design the  test based on $\trace(\M^\sir_{\calf\cup j})-\trace(\M^\sir_\calf)$.
The idea of using trace difference is similar to the extra sums of squares test in the classical
 multiple linear regression setting. Zhong et al. (2012) suggest a related test in the COP algorithm.
  Given the structural dimension $q$, denote the largest $q$ eigenvalues of $\M_{\calf\cup j}^\sir$ as $\lambda^{(k)}_{\calf\cup j}$, and the largest
$q$ eigenvalues of $\M_{\calf}^\sir$ as $\lambda^{(k)}_{\calf}$, $k=1,\ldots,q$. COP is based on the key quantity
$\sum_{k=1}^q (1-\lambda^{(k)}_{\calf\cup j})^{-1}(\lambda^{(k)}_{\calf\cup j}-\lambda^{(k)}_{\calf})$.
%Without the scaling factor $(1-\lambda^{(k)}_{\calf\cup j})^{-1}$,
%and assuming $\M_{\calf\cup j}^\sir$ and $\M_{\calf}^\sir$ both have rank $d$,
The COP test reduces to the trace test with
%$\sum_{k=1}^d (\lambda^{(k)}_{\calf\cup j}-\lambda^{(k)}_{\calf})$, or
$\trace(\M^\sir_{\calf\cup j})-\trace(\M^\sir_\calf)$ if we drop the
scaling factor $(1-\lambda^{(k)}_{\calf\cup j})^{-1}$ and
assume  both $\M_{\calf\cup j}^\sir$ and $\M_{\calf}^\sir$ have rank $q$. Compared with the COP test,
the SIR-based trace test does not involve estimating $q$. While COP can not be easily extended to dimension reduction methods other than SIR,
the trace test is a versatile framework and can be combined with methods other than SIR, as we will see next.

%    Denote $\var(\X)=\sig$, $R_h=I(\Y\in J_h)$,
%$p_h=E(R_h)$, and $\U_h=E(\X|\Y \in J_h)$. The
% kernel matrix for SIR is
%\begin{align}
%\label{eq:sir}
%\M^\sir=\halfinvsig\left(\sum_{h=1}^H p_h  \U_h \U_h^T\right)\halfinvsig.
%\end{align}
%Let $\V_h=E(\X\X^T|\Y \in J_h)$. The
% kernel matrix for SAVE can be rewritten as
% \begin{align}
%\label{eq:save}
%\M^\save=\sum_{h=1}^H p_h\{\sig^{-1/2}(\sig-\V_h+\U_h \U_h^T)\sig^{-1/2}\}^2.
%\end{align}
%

\csubsection{SAVE-based and directional regression-based trace tests}
\label{sec:save}

%
%SAVE  is an important extension of SIR. We denote  $\M^\save= E
%\left\{ \I_p-\var(\Z |  \Y\in J_h )\right\}^2$ as the
% kernel matrix for the discretized SAVE.
% Cook and Weisberg (1991)
%demonstrate that $\spn(\M^\save)$ $\subseteq \spc_{\Y|\Z}$ under
%the LCM assumption (\ref{eq:lcm}) and the
% constant conditional variance (CCV) assumption that
%\begin{align}
%\label{eq:ccv} \var(\X | \bfbeta^T \X) \mbox{ is nonrandom}.
%\end{align}
Note that $\M^\save$ can be rewritten as $\sum_{h=1}^H p_h
\{\sig^{-1/2}(\sig-\V_h+\U_h \U_h^T)\sig^{-1/2}\}^2$ with
$\V_h=E(\X\X^T|\Y \in J_h)$. Denote  $\V_{\calf, h}=E(\xf \xf^T|\Y
\in J_h)$ and define
$$ \M^\save_{\calf}=\sum_{h=1}^H p_h \{\sig_{\calf}^{-1/2}(\sig_{\calf}-\V_{\calf, h}+\U_{\calf, h}\U_{\calf, h}^T)\sig_{\calf}^{-1/2}\}^2.$$
For the purpose of sufficient dimension reduction, it is well-known that SAVE requires the
CCV assumption (\ref{eq:ccv}) in addition to the LCM assumption (\ref{eq:lcm}) required by SIR. In a parallel fashion, our SAVE-based trace test relies on the following
subset CCV assumption together with the subset LCM assumption (\ref{eq:subset lcm}),
 \begin{align}
\label{eq:subset ccv} \var(\x_j | \xf) \mbox{ is nonrandom} \mbox{
for any }\calf\subset \cali \mbox { and } j\in\calf^c.
\end{align}
Assumptions (\ref{eq:lcm}) and (\ref{eq:ccv}) are common in the sufficient dimension reduction literature.
Meanwhile, assumptions (\ref{eq:subset lcm}) and (\ref{eq:subset ccv}) have been used in Zhong et al. (2012) and Jiang and Liu (2013) for SIR based model-free variable selection.
All four conditions are satisfied when $\X$ is normal.
 The next theorem states the principle of the SAVE-based trace test.
\begin{theorem}\label{prop:save}
%Denote $\zeta_{j|\mathcal{F},h}=E(\gamma_{j|\mathcal{F}}^2|\Y\in J_h)$ and
%$\bfphi_{j|\mathcal{F},h}=\sig_{\calf}^{-1/2}\left\{ \U_{\calf, h}\gamma_{j|\mathcal{F},h}\right.$ $\left.-E(\xf\gamma_{j|\mathcal{F}}|\Y
%\in J_h)\right\}$.
 Assume the
subset LCM assumption (\ref{eq:subset lcm}) and the subset CCV assumption (\ref{eq:subset ccv}) hold true. Then  for
$\calf\subset \cali$ and $j\in \calf^c$, we have
\begin{enumerate}
\item  $\trace(\M^\save_{\calf\cup j})-\trace(\M^\save_\calf)=\sum_{h=1}^H p_h
\{(1-\zeta_{j|\mathcal{F},h}+\gamma_{j|\mathcal{F},h}^2)^2+2\bfphi_{j|\mathcal{F},h}^T\bfphi_{j|\mathcal{F},h}\}$,
where $\bfphi_{j|\mathcal{F},h}=\sig_{\calf}^{-1/2}\left\{
\U_{\calf, h}\gamma_{j|\mathcal{F},h}\right.$
$\left.-E(\xf\gamma_{j|\mathcal{F}}|\Y \in J_h)\right\}$ and
$\zeta_{j|\mathcal{F},h}=E(\gamma_{j|\mathcal{F}}^2|\Y\in J_h)$.
\item
%If the following
%subset CCV condition holds true in addition,
% \begin{align}
%\label{eq:subset ccv} \var(\x_j | \xf) \mbox{ is nonrandom} \mbox{
%for any }\calf\subset \cali \mbox { and } j\in\calf^c,
%\end{align}
 %then
   $\trace(\M^\save_{\calf\cup j})-\trace(\M^\save_\calf)=0$ given that
   %$\Y\indep \x_j| \xf$ and
   $\cala\subseteq\calf$.
\end{enumerate}
\end{theorem}

%Note that conditions (\ref{eq:subset lcm}) and (\ref{eq:subset ccv}) have been proposed in
%Zhong et al. (2012) for the COP algorithm. The subset LCM condition (\ref{eq:subset lcm}) is
%parallel to the LCM condition (\ref{eq:lcm}), while the subset CCV
%condition (\ref{eq:subset ccv}) is parallel to the CCV condition
%(\ref{eq:ccv}).
%  All four conditions are satisfied if $\X$ is
%multivariate normal.

%%\csubsection{Directional regression-based trace test}
%%\label{sec:dr}
%
%%Directional regression  is a popular sufficient dimension reduction
%%method which implicitly combines the strength of SIR and SAVE.
%The kernel
%matrix of directional regression can be rewritten as
%\begin{align*}
%&\M^\dr= 2\sum_{h=1}^H p_h (\sig^{-1/2} \V_h
%\sig^{-1/2})^2+2\left(\sum_{h=1}^H p_h
%\sig^{-1/2}\U_h \U_h^T \sig^{-1/2}\right)^2\\
%&\hspace{.2in}+2\left(\sum_{h=1}^H p_h \U_h^T \sig^{-1} \U_h \right)
%\left(\sum_{h=1}^H  p_h \sig^{-1/2} \U_h \U_h^T \sig^{-1/2}
%\right)-2\I_p.
%\end{align*}
%Li and Wang (2007) prove that $\spn(\M^\dr)\subseteq \spc_{\Y|\Z}$
%under conditions  (\ref{eq:lcm}) and (\ref{eq:ccv}).  We mimic
%$\M^\dr$ and define

 For directional regression-based trace test, define
\begin{align*}
&\M^\dr_{\calf}= 2\sum_{h=1}^H p_h (\sig^{-1/2}_{\calf} \V_{\calf,
h} \sig^{-1/2}_{\calf})^2+2\left(\sum_{h=1}^H p_h
\sig^{-1/2}_{\calf}\U_{\calf, h} \U_{\calf, h}^T \sig^{-1/2}_{\calf}\right)^2\\
&\hspace{.2in}+2\left(\sum_{h=1}^H p_h \U_{\calf, h}^T
\sig^{-1}_{\calf} \U_{\calf, h} \right) \left(\sum_{h=1}^H  p_h
\sig^{-1/2}_{\calf} \U_{\calf, h} \U_{\calf, h}^T
\sig^{-1/2}_{\calf} \right)-2\I_{|\mathcal{F}|},
\end{align*}
where $|\mathcal{F}|$ denotes the cardinality of $\calf$.
The directional regression-based trace test relies on the next
theorem.

\begin{theorem}\label{prop:dr}
 Assume the
subset LCM assumption (\ref{eq:subset lcm}) and the subset CCV assumption (\ref{eq:subset ccv}) hold true. Then  for
$\calf\subset \cali$ and $j\in \calf^c$, we have
\begin{enumerate}
\item  $\trace(\M^\dr_{\calf\cup j})-\trace(\M^\dr_\calf)=2\sum_{h=1}^H
p_h\left((1-\zeta_{j|\mathcal{F},h})^2+
2\bfnu_{j|\mathcal{F},h}^T\bfnu_{j|\mathcal{F},h}\right)+4\varrho_{j|\mathcal{F}}^2$
$+4\bfiota_{j|\mathcal{F}}^T
\bfiota_{j|\mathcal{F}}+4\kappa_{\mathcal{F}}\varrho_{j|\mathcal{F}}$,
where
$\bfnu_{j|\mathcal{F},h}=\sig^{-1/2}_{\mathcal{F}}E(\xf\gamma_{j|\mathcal{F}}|\Y
\in J_h)$,
$\bfiota_{j|\mathcal{F},h}=\sig^{-1/2}_{\mathcal{F}}\U_{\calf,h}\gamma_{j|\mathcal{F},h}$,
$\bfiota_{j|\mathcal{F}}=\sum_{h=1}^H p_h
\bfiota_{j|\mathcal{F},h}$, $\varrho_{j|\mathcal{F}}=\sum_{h=1}^H
p_h \gamma^2_{j|\mathcal{F},h}$, and
$\kappa_{\mathcal{F}}=\sum_{h=1}^H p_h \U_{\calf,h}^T
\sig^{-1}_{\mathcal{F}} \U_{\calf,h}$.
\item
%  If the
%subset CCV condition (\ref{eq:subset ccv}) holds true in addition,
%then
$\trace(\M^\dr_{\calf\cup j})-\trace(\M^\dr_\calf)=0$ given
that
%$\Y\indep \x_j| \xf$ and
$\cala\subseteq\calf$.
\end{enumerate}
\end{theorem}

Theorems \ref{prop:save} and \ref{prop:dr} demonstrate that the trace test can be a general framework. Unlike
COP, trace tests do not require estimation of the structural dimension $q$, and they can be combined with sufficient dimension reduction methods  other than SIR.

%\noindent The proofs for Theorems \ref{prop:sir},
%\ref{prop:save} and \ref{prop:dr} are provided in the Appendix \ref{pf sec2}.

\newpage

\csection{Asymptotic distributions of the trace test statistics}

  Given an i.i.d. sample $(\X_i, \Y_i)$, $i=1,\ldots,n$, we develop the asymptotic distribution of the
  sample level trace test statistics.
The asymptotic results in this section
%Theorems \ref{theorem:sir},
%\ref{theorem:save} and \ref{theorem:dr}
are developed with fixed
$|\calf|$ when $n$ goes to infinity.
For the SIR-based test,
% first
%, and the tests based on SAVE
%and directional regression can be developed in a similar fashion.
we have
\begin{theorem}\label{theorem:sir}
 Suppose $\X$
has finite fourth order moment, and the subset LCM assumption
(\ref{eq:subset lcm}) holds true.
%Define trace-test statistic
%$T^\sir_{j|\mathcal{F}}=n\left\{\trace(\hat{\M}^\sir_{\calf\cup
%j})-\trace(\hat{\M}^\sir_\calf) \right\}$.
 Then under
$H_0: \Y \indep \x_j| \xf$, $j\in \calf^c$, we have
\begin{align*}
T^\sir_{j|\mathcal{F}}\longrightarrow \sum_{k=1}^H
\omega_{j|\mathcal{F},k}^{\sir}\chi^2_1, \mbox{ where }T^\sir_{j|\mathcal{F}}=n\left\{\trace(\hat{\M}^\sir_{\calf\cup
j})-\trace(\hat{\M}^\sir_\calf) \right\}.
\end{align*}
Here
$\omega_{j|\mathcal{F},1}^{\sir}\ge\ldots\ge\omega_{j|\mathcal{F},H}^{\sir}$
are the eigenvalues of ${\bf \Omega}^\sir_{j|\mathcal{F}}$ defined
in the Appendix.
\end{theorem}

%follow Theorem \ref{prop:sir} and define test
%statistic
%\begin{align}
%\label{test sir}
%T^\sir_{j|\mathcal{F}}=n\left\{\trace(\hat{\M}^\sir_{\calf\cup
%j})-\trace(\hat{\M}^\sir_\calf) \right\}.
%%=n\sum_{h=1}^H \hat p_h
%%\hat\gamma_{j|\mathcal{F},h}^2.
%\end{align}
%The asymptotic distribution of $T^\sir_{j|\mathcal{F}}$ is provided next.

The test statistic $T^\sir_{j|\mathcal{F}}$
can be calculated as $n\sum_{h=1}^H \hat p_h
\hat\gamma_{j|\mathcal{F},h}^2$, where
 $\hat p_h$ and $\hat\gamma_{j|\mathcal{F},h}$ are sample counterparts of $p_h$ and
$\gamma_{j|\mathcal{F},h}$ defined in Theorem \ref{prop:sir}.
Since we assume $E(\X)=\0$ for the population level development,
$\hat\gamma_{j|\mathcal{F},h}$ is calculated based on
centered predictors.  Let $\tilde \X_i=(\tilde \x_{i1},\ldots,\tilde
\x_{ip})^T=\X_i-\sum_{i=1}^n \X_i/n$ be the centered version of
$\X_i$.
 Denote
$R_{i,h}=I(\Y_i\in J_h)$,
  $\hat p_h=\sum_{i=1}^n R_{i,h}/n$, $\tilde \X_{i(\calf)}=\{\tilde \x_{ij}: j\in \calf\}$, and
   $\hat{\U}_{\calf,h}=\sum_{i=1}^n \tilde \X_{i(\calf)} R_{i,h}/(n\hat p_h)$.
 Let $E_n(\tilde \x_j|\txf)$ be the sample estimator of $E(\tilde \x_j|\txf)$. Further denote
    $\hat\x_{ij|\mathcal{F}}=\tilde\x_{ij}-E_n(\tilde\x_j|\txf)$,
$\hat\sigma_{j|\mathcal{F}}^2=\sum_{i=1}^n
\hat\x_{ij|\mathcal{F}}^2/n-  (\sum_{i=1}^n
\hat\x_{ij|\mathcal{F}})^2/n^2 $,
and $\hat\gamma_{ij|\mathcal{F}}=\hat\x_{ij|\mathcal{F}}/\hat\sigma_{j|\mathcal{F}}$. Then we
have
%estimate $\gamma_{j|\mathcal{F},h}$ by
 $\hat\gamma_{j|\mathcal{F},h}=\sum_{i=1}^n
\hat\gamma_{ij|\mathcal{F}} R_{i,h}/(n\hat p_h)$.
   %Instead of inverting the $p\times p$ matrix
%  $\sig$, it only involves inverting the
% $|\calf|\times |\calf|$ matrix $\sig_{\mathcal{F}}$ and the $(|\calf|+1)\times (|\calf|+1)$ matrix $\sig_{\mathcal{F}\cup j}$.

The next two Theorems provide the asymptotic distribution for the SAVE-based and directional regression-based trace test statistics
respectively.
\begin{theorem}\label{theorem:save}
 Suppose $\X$
has finite fourth order moment. Assume the subset LCM assumption
(\ref{eq:subset lcm}) and the subset CCV assumption (\ref{eq:subset
ccv}) hold true. Then under $H_0: \Y \indep
\x_j| \xf$, $j\in \calf^c$, we have
\begin{align*}
T^\save_{j|\mathcal{F}}\longrightarrow \sum_{k=1}^{(|\mathcal
F|+1)H} \omega_{j|\mathcal{F},k}^\save\chi^2_1,  \mbox{ where }
T^\save_{j|\mathcal{F}}=n\left\{\trace(\hat{\M}^\save_{\calf\cup j})-\trace(\hat{\M}^\save_\calf)\right\}.
\end{align*}
Here
$\omega_{j|\mathcal{F},1}^\save\ge\ldots\ge\omega_{j|\mathcal{F},(|\mathcal
F|+1)H}^\save$ are the eigenvalues of ${\bf
\Omega}^\save_{j|\mathcal{F}}$ defined in the Appendix.
\end{theorem}

%We discuss the SAVE-based trace test statistic and its asymptotic
%distribution next. Define
%\begin{align}
%\label{test save}
%\begin{split}
%T^\save_{j|\mathcal{F}}&=n\left\{\trace(\hat{\M}^\save_{\calf\cup j})-\trace(\hat{\M}^\save_\calf) \right\}\\
%&=n\sum_{h=1}^H \hat p_h
%\{(1-\hat\zeta_{j|\mathcal{F},h}+\hat\gamma_{j|\mathcal{F},h}^2)^2+
%2\hat{\bfphi}^T_{j|\mathcal{F},h}\hat\bfphi_{j|\mathcal{F},h}\}.
%\end{split}
%\end{align}
%We then have

%
%For the directional regression-based trace test, define
%\begin{align}
%\label{test dr}
%\begin{split}
%T^\dr_{j|\mathcal{F}}&=n\left\{\trace(\hat{\M}^\dr_{\calf\cup
%j})-\trace(\hat{\M}^\dr_\calf) \right\}
%=4n\hat\varrho_{j|\mathcal{F}}^2+4n\hat\bfiota_{j|\mathcal{F}}^T
%\hat\bfiota_{j|\mathcal{F}}\\
%&\hspace{.1in}+4n\hat\kappa_{\mathcal{F}}\hat\varrho_{j|\mathcal{F}}+
%2n \sum_{h=1}^H \hat p_h\left((1-\hat\zeta_{j|\mathcal{F},h})^2  +
%2\hat\bfnu_{j|\mathcal{F},h}^T\hat\bfnu_{j|\mathcal{F},h}\right).
%\end{split}
%\end{align}
%The result is stated below.

\begin{theorem}\label{theorem:dr}
 Suppose $\X$
has finite fourth order moment. Assume the subset LCM assumption
(\ref{eq:subset lcm}) and the subset CCV assumption (\ref{eq:subset
ccv}) hold true. Then under $H_0: \Y \indep
\x_j| \xf$, $j\in \calf^c$, we have
\begin{align*}
T^\dr_{j|\mathcal{F}}\longrightarrow \sum_{k=1}^{2 |\mathcal
F|(H+1)} \omega_{j|\mathcal{F},k}^\dr\chi^2_1,  \mbox{ where }
T^\dr_{j|\mathcal{F}}=n\left\{\trace(\hat{\M}^\dr_{\calf\cup
j})-\trace(\hat{\M}^\dr_\calf) \right\}.
\end{align*}
Here
$\omega_{j|\mathcal{F},1}^\dr\ge\ldots\ge\omega_{j|\mathcal{F},2
|\mathcal F|(H+1)}^\dr$ are the eigenvalues of ${\bf
\Omega}^\dr_{j|\mathcal{F}}$ defined in the Appendix.
\end{theorem}

From Theorem \ref{prop:save}, we know $T^\save_{j|\mathcal{F}}$ can be calculated as $n\sum_{h=1}^H \hat p_h
\{(1-\hat\zeta_{j|\mathcal{F},h}+\hat\gamma_{j|\mathcal{F},h}^2)^2+
2\hat{\bfphi}^T_{j|\mathcal{F},h}\hat\bfphi_{j|\mathcal{F},h}\}$. From Theorem \ref{prop:dr}, we know $T^\dr_{j|\mathcal{F}}$ can be calculated as $4n\hat\varrho_{j|\mathcal{F}}^2+4n\hat\bfiota_{j|\mathcal{F}}^T
\hat\bfiota_{j|\mathcal{F}}+4n\hat\kappa_{\mathcal{F}}\hat\varrho_{j|\mathcal{F}}+
2n \sum_{h=1}^H \hat p_h\left((1-\hat\zeta_{j|\mathcal{F},h})^2  +
2\hat\bfnu_{j|\mathcal{F},h}^T\hat\bfnu_{j|\mathcal{F},h}\right)$.
To approximate the asymptotic distribution under $H_0$, we use
 estimated weights $\hat\omega_{j|\mathcal{F},k}^\sir$, $\hat\omega_{j|\mathcal{F},k}^\save$ and $\hat\omega_{j|\mathcal{F},k}^\dr$.
The detailed forms of these sample estimators
%$\hat\zeta_{j|\mathcal{F},h}$, $\hat\bfphi_{j|\mathcal{F},h}$, $\hat\bfnu_{j|\mathcal{F},h}$,
%$\hat\varrho_{j|\mathcal{F}}$, $\hat\bfiota_{j|\mathcal{F}}$ and
%$\hat\kappa_{\mathcal{F}}$
are provided in the
Appendix.
%\ref{pf sec3}

%Their population counterparts have been defined in Theorem
%\ref{prop:save} and Theorem \ref{prop:dr} respectively.
%At the sample level,
%, which are defined in the Appendix.

\csection{The stepwise trace pursuit algorithm}\label{sec:algorithm}

We provide the stepwise trace pursuit (STP) algorithm and its
selection consistency property in this section. For the ease of
presentation, the following stepwise algorithm is based on the SIR-based trace test.
The STP algorithms for SAVE and directional regression can be defined in a parallel fashion.

\begin{enumerate}
\item[(a)]  {\it Initialization}. Set the initial working set  to be $\calf=\varnothing$.
\item[(b)]  {\it Forward addition}.
Find index
$a_\calf$ such that
\begin{align}
\label{eq:af}
a_\calf=\underset{j\in \calf^c}{\mbox{arg max
}}  \trace\big(\hat
\M^\sir_{{\mathcal{F}}\cup j}\big).
\end{align}
If $T^{\textup{SIR}}_{a_\calf|\mathcal{F}}=n\{\trace\big(\hat
\M^\sir_{{\mathcal{F}}\cup a_\calf}\big) -\trace\big(\hat
\M^\sir_{{\mathcal{F}}}\big)\}>
   \bar c^\sir$, update
$\calf$ to be $\mathcal{F}\cup a_\calf$.
\item[(c)]  {\it Backward deletion}.
Find index $d_\calf$
such that
\begin{align}
\label{eq:df}
d_\calf=\underset{j \in \calf}{\mbox{arg max }}
\trace\big(\hat
\M^\sir_{{\mathcal{F}}\backslash j}\big).
\end{align}
%$d_\calf=\underset{j \in \calf}{\mbox{arg min }}
%\trace\big(\hat
%\M^\sir_{{\mathcal{F}}\backslash j}\big)$.
 If $T^{\textup{SIR}}_{d_\calf|\{{\mathcal{F}}\backslash
{d_\calf}\}}=n\{\trace\big(\hat
\M^\sir_{{\mathcal{F}}}\big)-\trace\big(\hat
\M^\sir_{{\mathcal{F}}\backslash {d_\calf}}\big)\}<
\underline{c}^{\textup{SIR}}$, update $\calf$ to be $\mathcal{F}\backslash d_\calf$.
\item[(d)]   Repeat steps (b) and (c) until no predictors can be added
or deleted.
\end{enumerate}

%The following proposition is helpful for us to study the selection consistency property of the stepwise
%SIR algorithm.

% We
%provide some additional insight about the key quantity $\trace(\M^\sir_{\calf\cup j})-\trace(\M^\sir_\calf)$ before we study the selection consistency property of the SIR-based STP algorithm.
%Recall that $\cala$ is the true set of significant
% predictors, $q$ denotes the structural dimension of the central space
%$\cals_{\Y|\X}$, and $\M^\sir=\var \left\{ E(\Z | \Y
%)\right\}$ has $q$ nonzero eigenvalues
%$\lambda_1\geq\ldots\geq\lambda_q$ with $\bfeta_1,\ldots, \bfeta_q$
%as the corresponding eigenvectors. Denote $\bfbeta_i=\sig^{-1/2}\bfeta_i=(\beta_{i,1},\ldots,\beta_{i,p})^T$ for $i=1,\ldots,q$, and let $\bfbeta_{\min}=\min_{1\le i\le q, j\in\cala,\beta_{i,j}\neq 0}|\beta_{i,j}|$.

 We
provide some additional insight about the key quantity $\trace(\M^\sir_{\calf\cup j})-\trace(\M^\sir_\calf)$ before we study the selection consistency property of the SIR-based STP algorithm.
Recall that $\cala$ is the true set of significant
 predictors, $q$ denotes the structural dimension of the central space
$\cals_{\Y|\X}$, and $\M^\sir=\var \left\{ E(\Z | \Y\in J_h
)\right\}$ has $q$ nonzero eigenvalues
$\lambda_1\geq\ldots\geq\lambda_q$ with $\bfeta_1,\ldots, \bfeta_q$
as the corresponding eigenvectors. Denote $\bfbeta_i=\sig^{-1/2}\bfeta_i=(\beta_{i,1},\ldots,\beta_{i,p})^T$ for $i=1,\ldots,q$, and let $\bfbeta_{\min}=\min_{j\in\cala}\sqrt{\sum_{i=1}^q \beta^2_{i,j}}$.

\begin{proposition}
\label{prop:max trace change} Assume $\spn(\bfbeta_1,\ldots,\bfbeta_q)=\cals_{\Y|\X}$ and the subset LCM assumption
(\ref{eq:subset lcm}) holds true.   Then
  for any $\calf$ such that $\calf^c\cap \cala \neq \varnothing$, we have
\begin{align*}
\max_{j \in \calf^c\cap \cala} \{ \trace(\M^\sir_{\calf\cup j})-\trace(\M^\sir_\calf)\} \ge \lambda_q\lambda^{-1}_{\max}(\sig) \lambda^{2}_{\min}(\sig) \bfbeta_{\min}^2,
\end{align*}
where $\lambda_{\max}(\sig)$ and $\lambda_{\min}(\sig)$ are the largest and the smallest eigenvalues of $\sig$ respectively.
\end{proposition}
We have seen in Theorem \ref{prop:sir} that $\trace(\M^\sir_{\calf\cup j})-\trace(\M^\sir_\calf)=0$ when $\cala\subseteq\calf$. Proposition \ref{prop:max trace change} implies that when $\cala$ does not belong to $\calf$, the maximum of $\trace(\M^\sir_{\calf\cup j})-\trace(\M^\sir_\calf)$ over $j \in \calf^c\cap \cala$ is greater than $0$. In the linear regression setting, $\bfbeta_{\min}$ can be explained as the minimum signal strength, and
 it is common to assume that $\bfbeta_{\min}$ does not decrease to $0$ too fast when
$n$ goes to infinity. This motivates us to assume that there exist $\varsigma>0$
and  $0<\xi_{\min}<1/2$ such that
\begin{align}
\label{eq:sep} \min_{\calf:\calf^c\cap\cala\neq \varnothing} \max_{j \in \calf^c\cap \cala} \{ \trace(\M^\sir_{\calf\cup j})-\trace(\M^\sir_\calf)\}>\varsigma n^{-\xi_{\min}}.
\end{align}

\begin{theorem}\label{theorem:selection}
 Suppose $\X$
has finite fourth order moment,  condition
(\ref{eq:subset lcm}) and condition (\ref{eq:sep}) hold true.
 \begin{enumerate}
\item If  we set $0<\bar c^\sir<\varsigma n^{1-\xi_{\min}}/2$, then as $n\rightarrow\infty$,
$$Pr(\min_{\calf:\calf^c\cap\cala\neq \varnothing} \max_{j \in \calf^c\cap \cala} T^{\textup{SIR}}_{j|\mathcal{F}}>
   \bar c^\sir )\rightarrow 1.$$
   \item If we set $\underline{c}^{\textup{SIR}}>C n^{1-\xi_{\min}}$ for any $C>0$, then as $n\rightarrow\infty$,
    $$Pr(\max_{\calf:\calf^c\cap\cala= \varnothing} \min_{j \in \calf} T^{\textup{SIR}}_{j|\{{\mathcal{F}}\backslash j\}}<
\underline{c}^{\textup{SIR}} )\rightarrow 1.$$
\end{enumerate}
\end{theorem}
Part~1 of Theorem \ref{theorem:selection} implies that the addition
step will not stop until all  significant predictors are selected. Part~2 implies that the deletion step of the algorithm will not stop if
the current selection includes any insignificant relevant predictors. Together,
they guarantee the selection consistency of the STP algorithm for SIR.
%Stepwise SAVE and stepwise directional regression can be
%developed in a parallel fashion.
To guarantee the selection
consistency of the
STP algorithms for SAVE and the directional regression,
condition (\ref{eq:sep})
in Theorem \ref{theorem:selection} has to be updated accordingly. We
leave the details to the Appendix. Note that the STP
algorithm is directly applicable even with $p>n$. All we need is
that $|\calf|<n$ for all iterations of the algorithm.

Condition (\ref{eq:sep}) relates closely to the concept of exhaustiveness in the literature of sufficient dimension reduction.
To fix the idea,
consider a toy example $Y=x_1^2$ and $x_i$ is i.i.d. $N(0,1)$ for $i=1,\ldots, p$. It's easy to see that $\trace(\M^\sir_{\calf\cup\{1\}})-\trace(\M^\sir_\calf)=0$ for any $\calf$, and condition (\ref{eq:sep}) is violated.
We expect the SIR-based STP algorithm to underfit
when condition (\ref{eq:sep}) is not satisfied.
%In the sufficient dimension reduction literature,
%to guarantee the efficacy of SIR and avoid the situations when $\spn(\M^\sir)\subset \spc_{\Y|\Z}$,
%From the proof of Proposition \ref{prop:max trace change}, we know
% condition (\ref{eq:sep}) implies the coverage condition (Cook, 2004) that $\spn(\bfbeta_1,\ldots,\bfbeta_q)=\cals_{\Y|\X}$.
%%{\bf While such coverage conditions might be violated for SIR and SAVE, it always hold for directional regression since directional regression is known to be
%%exhaustive in recovering the central subspace.}

\csection{The forward trace pursuit algorithm}\label{sec: forward trace pursuit}

To determine the index $a_\calf$ in the addition step of the STP
algorithm, we need to go over all possible candidate indices in
$\calf^c$ and compare  a total of $p-|\mathcal F|$ test statistics, which
may lead to overwhelming computation burden when $p$ is large.
%perform a total of $p-|\mathcal F|$ weighted $\chi^2$
%tests. $|\mathcal F|$ more  weighted $\chi^2$ tests are needed to
%determine the index $d_\calf$ in the deletion step of the STP
%algorithm. Together we need to carry out $p$ weighted $\chi^2$ tests
%within each iteration of the STP algorithm. Since the unknown
%parameters involved in the asymptotic null distributions of these
%tests have to be estimated from the sample, the computation of the
%STP algorithm becomes overwhelming when $p$ is large.
  This
motivates us to consider the forward trace pursuit (FTP) algorithm
as an initial screening step. The SIR-based FTP algorithm is
described as follows.

\begin{enumerate}
\item[(a)]  {\it Initialization}. Set ${\mathcal{S}}^{(0)}=\varnothing$.
\item[(b)]  {\it Forward addition}.
 For $k\ge1$, ${\mathcal{S}}^{(k-1)}$ is given at the beginning of the $k$th iteration.
For every $j\in \cali\backslash {\mathcal{S}}^{(k-1)}$, compute $\trace\big(\hat
\M^{\textup{SIR}}_{{\mathcal{S}}^{(k-1)}\cup j}\big)$, and find $a_k$
such that
\begin{align*}
a_k=\underset{j\in \cali\backslash {\mathcal{S}}^{(k-1)}}{\mbox{arg max
}}\trace\big(\hat \M^{\textup{SIR}}_{{\mathcal{S}}^{(k-1)}\cup
j}\big).
\end{align*}
\item[(c)]  {\it Solution path}. Repeat step (b) $n$ times, to get a sequence of $n$
nested candidate models. Denote the solution path as ${\mathcal{S}}
= \{\cals^{(k)} : 1 \le k \le n\}$, where ${\mathcal{S}}^{(k)} =
\{a_1,\ldots,a_k\}.$
\end{enumerate}

To study the theoretical property of forward trace pursuit based on SIR, we assume the following set of conditions.

\begin{itemize}
\item [(C1)] Assume that the predictor $\X$ is normally distributed.
\item [(C2)] Assume that there exist two positive constant $\tau_{\min}$ and $\tau_{\max}$, such that $\tau_{\min}<\lambda_{\min}(\sig)<\lambda_{\max}(\sig)<\tau_{\max}<\infty$. %Here $\lambda_{\min}(\sig)$ and $\lambda_{\max}(\sig)$ denote the smallest and the largest eigenvalues of $\sig$ respectively.
\item [(C3)] Assume condition (\ref{eq:sep}) holds true, and there exist constants $\xi$ and $\xi_0$, such that $\log p \le \varpi n^{\xi}$, $|\cala|\le \varpi n^{\xi_0}$, and $\xi+2\xi_{\min}+2\xi_0<1$.
\end{itemize}

(C1) and (C2) are commonly used conditions in high dimensional sparse covariance estimation and variable screening problems.
Wang (2009) assumed (C1) and (C2) to study the sure screening property of forward linear regression. Condition (C3) allows
both the predictor dimensionality $p$ and the number of significant predictors $|\cala|$ go to infinity  as $n$ goes to infinity. Note that (C1) implies condition (\ref{eq:subset lcm}).
Denote $[t]$ as the smallest integer no less than $t$.
We state the screening
consistency of the SIR-based FTP algorithm next.

\begin{theorem}\label{theo:sure screening} Assume conditions \textup{(C1)-(C3)} hold true. Then as $n \rightarrow \infty$ and $p \rightarrow \infty$, the solution path of the SIR-based FTP algorithm satisfies
\begin{align*}
Pr\left( \cala \subset \mathcal{S}^{([2H\varsigma^{-1}\varpi
n^{\xi_0+\xi_{\min}}])} \right) \rightarrow 1.
\end{align*}
\end{theorem}
%In the above theorem $C_0=2\tau_{\min}^{-2}\tau_{\max}\vartheta
%\vartheta_{\alpha}^{-4} C_{\alpha}^2$ is a constant independent of
%$n$. The constants $\tau_{\min}$, $\tau_{\max}$, $\vartheta$,
%$\vartheta_{\alpha}$, $C_{\alpha}$, $\xi_0$, and $\xi_{\min}$ are
%all defined through conditions \textup{(C1)-(C4)} in the Appendix \ref{pf sec5}.
Theorem~\ref{theo:sure screening} guarantees that the FTP based on SIR enjoys the sure screening property in a model free setting, which extends the
theoretical developments in Wang (2009). Moreover,
Theorem \ref{theo:sure screening}
implies that with $n$ going to
infinity, only a finite number of iterations is needed in the FTP
algorithm to recover the  set $\cala$ of true significant predictors if the dimension of
the true model is finite with $\xi_0=\xi_{\min}=0$. The proof of Theorem~\ref{theo:sure screening} requires delicate
asymptotic analysis and is relegated to the Appendix. The FTP algorithm
based on SAVE or the
directional regression can be developed parallel to SIR. Their
screening consistency properties are left for future
investigation.

 To choose one model from the entire solution
path ${\mathcal{S}} = \{\cals^{(k)} : 1 \le k \le n\}$, we follow  Chen and Chen (2008) and define the
modified  BIC criterion
\begin{align}\label{eq:BIC}
\textup{BIC}(\calf)=-\log\left\{\trace(\hat\M^\sir_{\calf})\right\}
+ n^{-1} |\calf| \left(\log n + 2\log p \right).
\end{align}
The candidate model ${\mathcal S}^{(\hat m)}$ is selected with $\hat
m=\argmin_{1\le k \le n} \textup{BIC}(\mathcal {S}^{(k)})$. The next
result states that the selected model enjoys the screening consistency property.
%${\mathcal S}^{(\hat m)}$ possesses
%the desirable screening consistency property.

\begin{theorem}\label{theo:BIC}
Assume conditions
\textup{(C1)-(C3)} hold true. Then as $n \rightarrow \infty$ and $p \rightarrow \infty$, $Pr\left( \cala \subset {\mathcal
S}^{(\hat m)} \right) \rightarrow 1$.
\end{theorem}
Theorem \ref{theo:BIC} suggests we use BIC to determine the model size of the FTP
algorithm in a data driven manner.
%Unlike existing  model-free variable screening methods such as the distance correlation (Li et al., 2012) and a nonparametric screening (Lin, et al. 2013), FTP with BIC does not need to pre-specify
%the model size and determines the model size in a data driven manner.
The hybrid trace pursuit (HTP) algorithm combines FTP as the initial
screening step and STP as the refined selection step. More
specifically, the SIR-based HTP algorithm works as follows.
\begin{enumerate}
\item[(a)]  Perform SIR-based FTP and get solution path ${\mathcal{S}}
= \{\cals^{(k)} : 1 \le k \le n\}$.
\item[(b)]  Based on BIC criterion (\ref{eq:BIC}), select ${\mathcal S}^{(\hat m)}$ with $\hat
m=\argmin_{1\le k \le n} \textup{BIC}(\mathcal {S}^{(k)})$.
\item[(c)] Perform the SIR-based STP, where the full index set $\cali=\{1,\ldots,p\}$ is updated to the screened index set
${\mathcal S}^{(\hat m)}.$
\end{enumerate}
The HTP algorithms for SAVE and the directional regression can be implemented similarly, and the details are omitted.

\hspace{.5in}

%\noindent Theorem \ref{theorem:selection} and Theorem
%\ref{theo:BIC} together imply that the SIR-based HTP algorithm could
%achieve the selection consistency with diverging $p$ if $|\cala|$ is fixed.
%The FTP and HTP algorithms
%for SAVE and the directional regression can be easily implemented along  the  lines of SIR.
%Following the proof of Theorem \ref{theo:BIC}, one can show that the BIC criterion based on SAVE or the directional regression will guarantee that all significant
%predictors are included when $p$ is fixed.
%Although we could
%use Proposition \ref{prop: ss prep} to facilitate the proof of
%theoretical property of SIR based FTP,
%Without the connection to linear model forward regression through Proposition \ref{prop: ss prep},

\csection{Numerical studies}

The proposed HTP algorithms are compared with existing model-free
variable selection methods in this section. The screening
performances of the FTP algorithms are evaluated as well.

\csubsection{Simulation studies}

 We consider the
following models:
\begin{align*}
%\textup{I}: \quad & Y= (x_1+x_2+x_{p-1}+x_p)^3+\epsilon,\\
\textup{I}: \quad & Y= \textup{sgn}(x_1+x_p) \exp(x_2+x_{p-1})+\epsilon,\\
%\textup{III}: \quad & Y= x_1^2+x_2^2+x_{p-1}^2+x_p^2+\epsilon,\\
\textup{II}: \quad & Y= 2x_1^2 x_p^2-2x_2^2 x_{p-1}^2+\epsilon,\\
\textup{III}: \quad & Y= x_1^4-x_p^4+3\exp(.8x_2+.6x_{p-1})+\epsilon.
%\\
%\textup{VI}: \quad & Y=2x_1+2x_p+1.5(x_2+x_{p-1})^2+\epsilon,\\
\end{align*}
Unless specified otherwise, we set ${\bf X}=(x_1,\ldots,x_p)^T$ to be multivariate normal with
$E(\X)=\0$ and $\var(\X)=\sig$, and $\epsilon\sim N(0,\sigma^2)$ is
independent of $\X$. The structural dimensions for Models I to III
are  respectively $q=2$, $4$ and $3$.
The  index set of significant predictors for all the three models
is $\cala=\{1,2,p-1,p\}$. In all the simulation studies, we set
$\sigma=.2$, the sample size $n=300$, the number of slices $H=4$. Consider
three settings of $p$: $p=10$ for small dimensionality, $p=100$ for
moderate dimensionality, and $p=1000$ for high dimensionality.
Denote the $(i,j)$th entry of $\sig$ as $\rho^{|i-j|}$, and in the simulations,
$\rho=0$ is with uncorrelated predictors  and $\rho=.5$ with correlated predictors.

When  the SIR-based STP algorithm described in Section
4 is implemented, the threshold values
$\bar{c}^\sir$ and
$\underline{c}^{\textup{SIR}}$ cannot be easily determined as they depend on unknown
rate $\xi_{\min}$ relative to the sample size.
Denote $D^{\textup{SIR}}_{j|\mathcal{F}}$ as the weighted $\chi^2$
distribution under $H_0: \Y \indep \x_j| \xf$ in Theorem \ref{theorem:sir}.
It is easier in practice to choose quantiles of $D^{\textup{SIR}}_{j|\mathcal{F}}$ as the threshold values
 for the test statistics
$T^{\textup{SIR}}_{a_\calf|\mathcal{F}}$ and
$T^{\textup{SIR}}_{d_\calf|\{{\mathcal{F}}\backslash {d_\calf}\}}$.
Recall the definitions of $a_\calf$ and $d_\calf$ in (\ref{eq:af}) and
(\ref{eq:df}).
%it is easier in practice to choose the threshold values for the
%$p$-values $p^{\textup{SIR}}_{a_\calf|\mathcal{F}}$ and
%$p^{\textup{SIR}}_{d_\calf|\{{\mathcal{F}}\backslash {d_\calf}\}}$,
%where $a_\calf$ and $d_\calf$ are defined in (\ref{eq:af}) and
%(\ref{eq:df}) respectively.
% Recall that
%$D^{\textup{SIR}}_{j|\mathcal{F}}$ denotes the weighted $\chi^2$
%distribution under $H_0$ in Theorem \ref{theorem:sir}.
 For the
forward addition step,
    $\calf$ is updated
to be $\mathcal{F}\cup a_\calf$ if
$T^{\textup{SIR}}_{a_\calf|\mathcal{F}}>D^{\textup{SIR}}_{\alpha,a_\calf|\mathcal{F}}$,
the $\alpha$th upper quantile of
$D^{\textup{SIR}}_{a_\calf|\mathcal{F}}$. Similarly, in the backward
deletion step, $\calf$ is updated to be $\mathcal{F}\backslash
d_\calf$ if $T^{\textup{SIR}}_{d_\calf|\{{\mathcal{F}}\backslash
{d_\calf}\}}<
D^{\textup{SIR}}_{\alpha,d_\calf|\{{\mathcal{F}}\backslash
{d_\calf}\}}$, the $\alpha$th upper quantile of
$D^{\textup{SIR}}_{d_\calf|\{{\mathcal{F}}\backslash {d_\calf}\}}$.
 Jiang and Liu (2013)
suggest trying $\alpha$ over the grid points in the interval $(0,1)$, and
determining the final $\alpha$ by the cross validation. For ease of implementation, we set the level
of $\alpha$ as $.1p^{-1}$ in all the simulation studies.
 We follow Bentler and Xie (2000) to approximate the
$\alpha$th upper quantile of a weighted $\chi^2$ distribution. Other approximations, such as Field (1993), Cook and Setodji (2003),
can be used as well. The STP algorithms based on SAVE and directional regression are carried out in a similar fashion.

%{\bf We follow Bentler and Xie (2000) to approximate the
%$100\alpha$th quantile of $\sum_{k=1}^H
%\hat\omega_{j|\mathcal{F},k}^{\sir}\chi^2_1$, where $\hat\omega_{j|\mathcal{F},k}^{\sir}$ is the estimated weight coefficient for $\omega_{j|\mathcal{F},k}^{\sir}$, see (\ref{eq:est Omega}) for more details.} Other approximations of sums of weighted $\chi^2$ distribution, such as Field (1993), Cook and Setodji (2003),
%can be used as well.

%suggesting the efficacy of the screening step via forward SIR.

We examine the performances of the HTP algorithms for variable selection in Tables \ref{table1-1} to \ref{table2}. The HTP algorithms that are based on SIR, SAVE and the directional regression are
denoted by HTP-SIR, HTP-SAVE and HTP-DR respectively.
  Based on the
$N=100$ repetitions, we report the underfitted count (UF), the
correctly fitted count (CF), the overfitted count (OF), and the
average model size (MS). Let $\widehat\cala_{(i)}$ be the estimated
active set in the $i$th repetition and define
\begin{align*}
&UF=\sum_{i=1}^{N} I(\cala \not\subseteq \widehat\cala_{(i)}), CF=\sum_{i=1}^{N} I(\cala = \widehat\cala_{(i)}),\\
&OF=\sum_{i=1}^{N} I(\cala \subset \widehat\cala_{(i)}), \mbox{ and
} MS=N^{-1}\sum_{i=1}^{N} |\widehat\cala_{(i)}|.
\end{align*}

The selection performance of Model I is summarized in Table \ref{table1-1}.
This model favors SIR as
$Y$ is monotone of the two linear combinations $x_1+x_p$ and
$x_2+x_{p-1}$.
HTP-SIR
works very well for this model, as condition (\ref{eq:sep}) is
satisfied here. The performance of HTP-SIR keeps up with diverging $p$, which
validates our theoretical finding in Theorem \ref{theo:BIC}.
 We know from the
sufficient dimension reduction literature that SAVE is not efficient when
predictors are linked to the response through monotone functions. We see that
 HTP-SAVE has very unstable
performances, which either underfits or overfits with
a large probability. HTP-DR
 performs similarly to HTP-SIR, and fits correctly with a dominant
probability.

\begin{table}
  \centering
%\resizebox{\textwidth}{!}{\begin{minipage}{\textwidth}
{\begin{minipage}{14cm}
\caption{Comparison among three HTP algorithms for Model I.
%Selection performances  based on $N=100$ repetitions are reported.
}
\label{table1-1}
\begin{tabular}{|c|c|c|c|c|c|c|c|c|c|c|}
%\multicolumn{10}{ c } { {\sc Table 1}: Model I results based on $N=100$ repetitions.}  \\
\hline & &    & \multicolumn{4}{c|} {$\rho=0$} & \multicolumn{4}{c|}
{$\rho=.5$} \\ \hline Model & Method &  $p$ &  UF  &  CF &  OF  &
MS &
UF  &  CF &  OF  &  MS          \\
\hline
          && $10$  &0 &100  &0  &4.00  &0  &100  &0  &4.00
\\ \cline {3-11}
& HTP-SIR  & $100$  &0  &100  &0  &4.00  &0  &100 &0  &4.00   \\
\cline {3-11}
          && $1000$  &0  &100  &0  &4.00  &0  &100  &0  &4.00  \\ \cline {2-11}
          && $10$  &9  &59  &32  &4.31  &4  &39 &57 &4.00 \\ \cline {3-11}
I &HTP-SAVE  & $100$  &32  &0  &68  &20.53  &46  &1 &53 &18.14  \\
\cline {3-11}
          && $1000$  &90  &0  &10  &18.93  &91  &0  &9  &15.59  \\ \cline {2-11}
          && $10$  &0  &98  &2  &4.02  &0  &99  &1  &4.01   \\ \cline {3-11}
 &HTP-DR  & $100$  &0  &95  &5  &4.07  &0  &93  &7  &4.08   \\
\cline {3-11}
          && $1000$  &0  &96  &4  &4.04  &0  &94 &6 &4.08  \\ \hline
%          \hline
%          && $10$  &100  &0  &0  & .31  &100  &0  &0  &.24  \\ \cline {3-11}
% &HTP-SIR  & $100$  &100  &0  &0  &.13  &100  &0  &0  &.08  \\ \cline {3-11}
%          && $1000$  &100  &0  &0  &.03  &100  &0  &0  &.03  \\ \cline {2-11}
%          && $10$  &2  &97  &1  &3.99  &2  &94  &4  &4.02 \\ \cline {3-11}
% II&HTP-SAVE  & $100$  &3  &53  &44  &4.63  &3  &50  &47 &4.71  \\ \cline {3-11}
%          && $1000$  &3  &48  &49  &4.79  &7  &41  &52  &4.95  \\ \cline {2-11}
%          && $10$  &3  &95  &2  &3.99  &3  &93  &4  &4.01  \\ \cline {3-11}
% &HTP-DR  & $100$  &2  &56  &42  &4.70  &3  &46  &51  &4.77  \\ \cline {3-11}
%          && $1000$  &7  &44  &49  &4.76  &7  &45  &48 &4.91  \\ \hline
%          \hline
%          && $10$  &100  &0  &0  &2.12  &100  &0  &0  &2.18  \\ \cline {3-11}
% &HTP-SIR  & $100$  &100  &0  &0  &2.53  &100  &0  &0  &2.52  \\ \cline {3-11}
%          && $1000$  &100  &0  &0  &3.82  &100  &0  &0  &2.03  \\ \cline {2-11}
%          && $10$  &7  &40  &53  &4.92  &9  &34  &57 &4.85 \\ \cline {3-11}
% III&HTP-SAVE  & $100$  &43  &1  &56  &6.58  &31  &2  &67 &6.29   \\ \cline {3-11}
%          && $1000$  &93  &0  &7  &9.26  &90  &6  &4  &3.27  \\ \cline {2-11}
%          && $10$  &0  &89  &11  &4.14  &1  &87  &12  &4.10 \\ \cline {3-11}
% &HTP-DR  & $100$   &3  &74  &22  &4.23  &2  &81  &17  &4.17  \\ \cline {3-11}
%          && $1000$  &6  &58  &36  &4.60  &7  &59  &34 &4.52  \\ \hline
\end{tabular}
  \end{minipage}}
\end{table}

\begin{table}
  \centering
%\resizebox{\textwidth}{!}{\begin{minipage}{\textwidth}
{\begin{minipage}{14cm}
\caption{Comparison among three HTP algorithms for Model II.
%Selection performances  based on $N=100$ repetitions are reported.
}
\label{table1-2}
\begin{tabular}{|c|c|c|c|c|c|c|c|c|c|c|}
%\multicolumn{10}{ c } { {\sc Table 1}: Model I results based on $N=100$ repetitions.}  \\
\hline & &    & \multicolumn{4}{c|} {$\rho=0$} & \multicolumn{4}{c|}
{$\rho=.5$} \\ \hline Model & Method &  $p$ &  UF  &  CF &  OF  &
MS &
UF  &  CF &  OF  &  MS          \\
%\hline
%          && $10$  &0 &100  &0  &4.00  &0  &100  &0  &4.00
%\\ \cline {3-11}
%& HTP-SIR  & $100$  &0  &100  &0  &4.00  &0  &100 &0  &4.00   \\
%\cline {3-11}
%          && $1000$  &0  &100  &0  &4.00  &0  &100  &0  &4.00  \\ \cline {2-11}
%          && $10$  &9  &59  &32  &4.31  &4  &39 &57 &4.00 \\ \cline {3-11}
%I &HTP-SAVE  & $100$  &32  &0  &68  &20.53  &46  &1 &53 &18.14  \\
%\cline {3-11}
%          && $1000$  &90  &0  &10  &18.93  &91  &0  &9  &15.59  \\ \cline {2-11}
%          && $10$  &0  &98  &2  &4.02  &0  &99  &1  &4.01   \\ \cline {3-11}
% &HTP-DR  & $100$  &0  &95  &5  &4.07  &0  &93  &7  &4.08   \\
%\cline {3-11}
%          && $1000$  &0  &96  &4  &4.04  &0  &94 &6 &4.08  \\ \hline
          \hline
          && $10$  &100  &0  &0  & .31  &100  &0  &0  &.24  \\ \cline {3-11}
 &HTP-SIR  & $100$  &100  &0  &0  &.13  &100  &0  &0  &.08  \\ \cline {3-11}
          && $1000$  &100  &0  &0  &.03  &100  &0  &0  &.03  \\ \cline {2-11}
          && $10$  &2  &97  &1  &3.99  &2  &94  &4  &4.02 \\ \cline {3-11}
 II&HTP-SAVE  & $100$  &3  &53  &44  &4.63  &3  &50  &47 &4.71  \\ \cline {3-11}
          && $1000$  &3  &48  &49  &4.79  &7  &41  &52  &4.95  \\ \cline {2-11}
          && $10$  &3  &95  &2  &3.99  &3  &93  &4  &4.01  \\ \cline {3-11}
 &HTP-DR  & $100$  &2  &56  &42  &4.70  &3  &46  &51  &4.77  \\ \cline {3-11}
          && $1000$  &7  &44  &49  &4.76  &7  &45  &48 &4.91  \\ \hline
%          \hline
%          && $10$  &100  &0  &0  &2.12  &100  &0  &0  &2.18  \\ \cline {3-11}
% &HTP-SIR  & $100$  &100  &0  &0  &2.53  &100  &0  &0  &2.52  \\ \cline {3-11}
%          && $1000$  &100  &0  &0  &3.82  &100  &0  &0  &2.03  \\ \cline {2-11}
%          && $10$  &7  &40  &53  &4.92  &9  &34  &57 &4.85 \\ \cline {3-11}
% III&HTP-SAVE  & $100$  &43  &1  &56  &6.58  &31  &2  &67 &6.29   \\ \cline {3-11}
%          && $1000$  &93  &0  &7  &9.26  &90  &6  &4  &3.27  \\ \cline {2-11}
%          && $10$  &0  &89  &11  &4.14  &1  &87  &12  &4.10 \\ \cline {3-11}
% &HTP-DR  & $100$   &3  &74  &22  &4.23  &2  &81  &17  &4.17  \\ \cline {3-11}
%          && $1000$  &6  &58  &36  &4.60  &7  &59  &34 &4.52  \\ \hline
\end{tabular}
  \end{minipage}}
\end{table}

Table \ref{table1-2} reports the performance of HTP methods for Model II, which favors SAVE as $Y$ depends on
quadratic functions $x_1^2,x_2^2,x_{p-1}^2$ and $x_p^2$.
 HTP-SAVE works reasonably well for this model. It correctly recovers $\cala$ with a dominant probability when $p=10$.
With probability close to one, HTP-SAVE either correctly
identifies or overfits
 $\cala$ when $p=100$ or $p=1000$. The average model size of HTP-SAVE for Model II is not much larger
than $4$, indicating very mild overfitting.
 From the average model size, we see that HTP-SIR
underfits and misses all four variables with high probability. This is as expected because condition (\ref{eq:sep}) is violated for this
model. HTP-DR
 has very similar performance to HTP-SAVE for this model.

%We first focus on HTP-SIR in Table \ref{table1} and have the following observations. HTP-SIR
%works very well for Model I. It underfits for both Models II and III. On average,
%HTP-SIR misses four variables for Model II, and misses two variables
%for Model III. This confirms our comment right after Theorem~
%\ref{theorem:selection}. Namely, the SIR-based stepwise algorithm
%achieves selection consistency  when condition (\ref{eq:sep}) is
%satisfied, and underfits when condition (\ref{eq:sep}) is violated.

%  that HTP-SIR can
% still be consistent with diverging $p$.
%From Table \ref{table1}, we see HTP-SIR works very well for Model I,
%and correctly identifies the active predictors across all settings.
%HTP-SIR almost always underfits for Models II and III. From the
%average model size, we see that HTP-SIR misses all four variables
%for Model II, and misses two variables on average for Model III.
%This is as expected, as HTP-SIR will miss the quadratic terms in
%Model II and the quartic terms in Model III.

The comparison of Model III is reported in Table \ref{table1-3}.
Model III
involves quartic functions $x_1^4$, $x_p^4$,
as well as a monotone function $3\exp(.8x_2+.6x_{p-1})$, and is thus favorable for the directional regression.
Both HTP-SIR  and
HTP-SAVE would fail.
HTP-SIR misses two variables on average, which should be $x_1$ and $x_p$ involved in the two quartic terms.
As we have seen in Table \ref{table1-1}, HTP-SAVE  either underfits or overfits with
a large probability due to the monotone function $3\exp(.8x_2+.6x_{p-1})$. HTP-DR still
enjoys good performance for Model III, and correctly recovers $\cala$ with a large probability.
The average model size is always close to $4$,
indicating a good overall fit.

%As the directional regression is designed to implicitly combine
%SIR and SAVE, we expect  HTP-DR to inherit the strength of both
%HTP-SIR and HTP-SAVE. It turns out that HTP-DR
% performs similarly to HTP-SIR for Model
%I, and similarly to HTP-SAVE for Model II.  HTP-DR still
%enjoys good performance for Model III, when both HTP-SIR  and
%HTP-SAVE would fail. Across all the three models, HTP-DR seldom
%underfits and correctly fits with a dominant
%probability. The average model size is always close to $4$,
%indicating a good overall fit.

\begin{table}
  \centering
%\resizebox{\textwidth}{!}{\begin{minipage}{\textwidth}
{\begin{minipage}{14cm}
\caption{Comparison among three HTP algorithms for Model III.
%Selection performances  based on $N=100$ repetitions are reported.
}
\label{table1-3}
\begin{tabular}{|c|c|c|c|c|c|c|c|c|c|c|}
%\multicolumn{10}{ c } { {\sc Table 1}: Model I results based on $N=100$ repetitions.}  \\
\hline & &    & \multicolumn{4}{c|} {$\rho=0$} & \multicolumn{4}{c|}
{$\rho=.5$} \\ \hline Model & Method &  $p$ &  UF  &  CF &  OF  &
MS &
UF  &  CF &  OF  &  MS          \\
%\hline
%          && $10$  &0 &100  &0  &4.00  &0  &100  &0  &4.00
%\\ \cline {3-11}
%& HTP-SIR  & $100$  &0  &100  &0  &4.00  &0  &100 &0  &4.00   \\
%\cline {3-11}
%          && $1000$  &0  &100  &0  &4.00  &0  &100  &0  &4.00  \\ \cline {2-11}
%          && $10$  &9  &59  &32  &4.31  &4  &39 &57 &4.00 \\ \cline {3-11}
%I &HTP-SAVE  & $100$  &32  &0  &68  &20.53  &46  &1 &53 &18.14  \\
%\cline {3-11}
%          && $1000$  &90  &0  &10  &18.93  &91  &0  &9  &15.59  \\ \cline {2-11}
%          && $10$  &0  &98  &2  &4.02  &0  &99  &1  &4.01   \\ \cline {3-11}
% &HTP-DR  & $100$  &0  &95  &5  &4.07  &0  &93  &7  &4.08   \\
%\cline {3-11}
%          && $1000$  &0  &96  &4  &4.04  &0  &94 &6 &4.08  \\ \hline
%          \hline
%          && $10$  &100  &0  &0  & .31  &100  &0  &0  &.24  \\ \cline {3-11}
% &HTP-SIR  & $100$  &100  &0  &0  &.13  &100  &0  &0  &.08  \\ \cline {3-11}
%          && $1000$  &100  &0  &0  &.03  &100  &0  &0  &.03  \\ \cline {2-11}
%          && $10$  &2  &97  &1  &3.99  &2  &94  &4  &4.02 \\ \cline {3-11}
% II&HTP-SAVE  & $100$  &3  &53  &44  &4.63  &3  &50  &47 &4.71  \\ \cline {3-11}
%          && $1000$  &3  &48  &49  &4.79  &7  &41  &52  &4.95  \\ \cline {2-11}
%          && $10$  &3  &95  &2  &3.99  &3  &93  &4  &4.01  \\ \cline {3-11}
% &HTP-DR  & $100$  &2  &56  &42  &4.70  &3  &46  &51  &4.77  \\ \cline {3-11}
%          && $1000$  &7  &44  &49  &4.76  &7  &45  &48 &4.91  \\ \hline
          \hline
          && $10$  &100  &0  &0  &2.07  &100  &0  &0  &2.23  \\ \cline {3-11}
 &HTP-SIR  & $100$  &100  &0  &0  &2.58  &100  &0  &0  &3.56  \\ \cline {3-11}
          && $1000$  &100  &0  &0  &6.34  &100  &0  &0  &6.56  \\ \cline {2-11}
          && $10$  &4  &33  &63  &5.14  &7  &45  &48 &4.61 \\ \cline {3-11}
 III&HTP-SAVE  & $100$  &47  &8  &45  &12.42  &36  &6 &58 &8.43   \\ \cline {3-11}
          && $1000$  &86  &0  &14  &21.76  &78  &2  &20  &16.86  \\ \cline {2-11}
          && $10$  &0  &91  &9  &4.11  &0  &98  &2  &4.02 \\ \cline {3-11}
 &HTP-DR  & $100$   &3  &83  &14  &4.13  &4  &79  &17  &4.16  \\ \cline {3-11}
          && $1000$  &4  &88  &8  &4.06  &5  &61  &34 &4.39  \\ \hline
\end{tabular}
  \end{minipage}}
\end{table}

\begin{table}
  \centering
%\resizebox{\textwidth}{!}{\begin{minipage}{\textwidth}
{\begin{minipage}{15.5cm}
\caption{Comparison among three HTP algorithms for Model III with  nonnormal $\X$. }
\label{table1-4}
\begin{tabular}{|c|c|c|c|c|c|c|c|c|c|c|c|c|}
%\multicolumn{10}{ c } { {\sc Table 1}: Model I results based on $N=100$ repetitions.}  \\
\hline $p=1000$ & \multicolumn{4}{c|} {case (i), Uniform} & \multicolumn{4}{c|}
{case (ii), Exponential}  & \multicolumn{4}{c|}
{case (iii), Geometric}\\ \hline Method &  UF  &  CF &  OF  &
MS &
UF  &  CF &  OF  &  MS  &
UF  &  CF &  OF  &  MS        \\ \hline
% COP & & & & & &　& & & & & & \\ \hline
% SIRI & & & & & &　& & & & & & \\ \hline
 HTP-SIR  &0  &88  &12  &4.12  &0  &99  &1  &4.01  &1  &98  &1  &4.00 \\ \hline
 HTP-SAVE  &0  &25  &75  &5.62  &7  &72  &19  &4.39  &27  &17  &56  &5.58 \\ \hline
 HTP-DR  &0  &93  &7  &4.08  &5  &81  &14  &4.06  &11  &86  &3  &3.82 \\ \hline
 %\hline
% SIS & & & & & &　& & & & & & \\ \hline
% DC-SIS & & & & & &　& & & & & & \\ \hline
% FTP-SIR  &0  &88  &12  &4.12  &0  &99  &1  &4.01  &1  &98  &1  &4.01 \\ \hline
% FTP-SAVE  &0  &0  &100  &23.76  &0  &0  &100  &14.92  &0  &0  &100  &13.85 \\ \hline
% FTP-DR  &0  &0  &100  &19.93  &0  &0  &100  &9.42  &0  &0  &100  &8.93 \\ \hline
\end{tabular}
  \end{minipage}}
\end{table}

To check the performances of the HTP algorithms for nonnormal predictors, consider
${\bf X}=(x_1,\ldots,x_p)^T$ with : case (i), $x_i\sim \textrm{Uniform}(1,2)$; case (ii),
$x_i\sim \textrm{Exponential}(1)$; case (iii), $x_i\sim \textrm{Geometric}(.5)$. In all three cases,
$x_i$ is independent of $x_j$ for $i\neq j$, $1\leq i, j\leq p$. We focus on Model III with $n=300$ and $p=1000$, and report the results in Table \ref{table1-4}.  We see that HTP-DR with nonnormal predictors has similar performance compared to its counterpart with normal predictors in Table \ref{table1-3}.  HTP-SAVE has unstable performance as before. The performance of HTP-SIR with nonnormal predictors actually has significant improvement over its counterpart with normal predictors in Table \ref{table1-3}. We have seen before that SIR-based method can not pick up quartic terms
$x_1^4$, $x_p^4$ involved in
Model III with $x_i\sim N(0,1)$. This happens because the symmetry of the $x_i$ distribution coincides with the symmetry of the link function $x_i^4$. Since the distribution of the nonnormal $x_i$ is no longer symmetric about $0$, HTP-SIR is able to select $x_1$ and $x_p$ in the quartic terms. We conclude from Table \ref{table1-4} that the proposed HTP algorithms are not sensitive to the normality assumption of the predictors.

\begin{table}
{\begin{center}
\caption{Comparison between COP, SIRI and HTP-DR. Selection
performances based on $p=1000$ and $N=100$ repetitions are
reported.} \label{table2}
\begin{tabular}{|c|c|c|c|c|c|c|c|c|c|}
\hline & &     \multicolumn{4}{c|} {$\rho=0$} & \multicolumn{4}{c|}
{$\rho=.5$} \\
\hline Model & Method &   UF  &  CF &  OF  & MS & UF  &  CF &  OF  &
MS          \\ \hline
          & COP  &0  &86  &14  &4.14  &0  &85  &15  &4.16 \\ \cline{2-10}
          I& SIRI &0  &66  &34  &4.46  &0  &86  &14  &4.19 \\ \cline{2-10}
          & HTP-DR   &0  &96  &4  &4.04 &0 &94  &6 &4.08  \\ \hline
          & COP  &100 &0  &0  &4.00  &100 &0  &0  &4.00 \\ \cline{2-10}
          II& SIRI &52  &38  &10  &3.79  &36  &45  &19  &4.05 \\ \cline {2-10}
          & HTP-DR   &7  &44  &49  &4.76  &7  &45  &48  &4.91\\ \hline
          & COP  &100 &0  &0  &3.09  &100 &0  &0  &3.15 \\ \cline{2-10}
          III& SIRI &1  &99  &0  &3.99  &2  &98  &0  &3.98 \\ \cline{2-10}
          & HTP-DR   &4  &88  &8  &4.06  &5  &61  &34 &4.39         \\ \hline
\end{tabular}
  \end{center}}
\end{table}

Next we focus on the challenging case of $p=1000$, and compare the
performances of HTP-DR with existing methods in Table \ref{table2}.
%based on $N=100$ repetitions.
Only COP (Zhong et al., 2012) and SIRI
(Jiang and Liu, 2013) are included in the comparison, as other methods
such as CISE (Chen et al., 2010) can not handle  $p>n$.
 The R codes for COP and SIRI are made available
by the respective authors. COP  works well for Model I, and
underfits for Models II and III. COP has similar performances to
HTP-SIR as both the methods are based on SIR. SIRI works well for Models I and III, and is likely to underfit for Model II.
We suspect the relatively large  structure dimension $q=4$ in Model II is a probable
cause for the deficiency of SIRI. HTP-DR completely avoids estimating the structure
dimension $q$, and has decent overall performance.

\begin{table}
{\begin{center}
\caption{Comparison between SIS, DC-SIS and FTP algorithms for
screening. Frequencies of cases including all active predictors are
reported based on $p=2000$ and $N=100$ repetitions.} \label{table3}
\begin{tabular}{|c|c|c|c|c|c|c|}
\hline &    \multicolumn{3}{c|} {$\rho=0$} & \multicolumn{3}{c|}
{$\rho=.5$} \\
\hline  Method &   Model I  &  Model II &  Model III    &   Model I
&  Model II &  Model III
\\ \hline
  SIS    & 15  &0  &3  & 97 & 1 & 23\\ \hline
 DC-SIS & 100  &100  &100  & 100 & 100 & 100\\ \hline
 FTP-SIR  & 100  &0  &0  & 100 & 0 & 0\\ \hline
 FTP-SAVE& 12  &97  &7 & 10 & 98 & 31\\ \hline
 FTP-DR    & 100  &97  &98 & 100 & 98 & 97\\ \hline
\end{tabular}
  \end{center}}
\end{table}

% siri rho=0 0.0100 0.9900 0.0000 3.9900
% rho=.5 1.000000e-02 9.800000e-01 1.000000e-02 4.000000e+00

To compare the screening performances of the FTP algorithms, we
report in Table \ref{table3} the frequencies in $N=100$ repetitions
when all the significant predictors are included after screening. The FTP
algorithms that are based on SIR, SAVE and the directional regression are denoted
by FTP-SIR, FTP-SAVE and FTP-DR respectively. The sure independence
screening (SIS) in Fan and Lv (2008) and the distance correlation-based SIS (DC-SIS) (Li et al., 2012) are also included. The model size of SIS and DC-SIS is set to be  $[n/\log n]$, while the model size of FTP is determined
by the  BIC criterion in  (\ref{eq:BIC}). SIS does not
work well
 as it is designed for linear models.
% DC-SIS has a decent overall performance, except that
%it may have trouble with higher order terms in Model III when the
%predictors are uncorrelated. This limitation is alleviated when
%the correlation is moderate among the predictors.
 FTP-SIR works well for Model I as the significant predictors appear in the monotone link functions.
FTP-SAVE works well for Model II as the significant predictors appear in
the quadratic link functions. FTP-DR performs similarly to the state of the art method DC-SIS,
and retains all active predictors with probability close to one
across all the three models. In addition, the BIC in (\ref{eq:BIC}) for FTP-DR
leads to average model size of 20, which is much smaller compared to
$[300/\log 300]=53$, the model size of DC-SIS.

\begin{table}
\caption{Classification results based on LDA for the leukemia data.} \label{table4}
\begin{tabular}{|c|c|c|c|c|c|}
\hline  Method &   COP  & SIRI &  HTP-SIR  &  HTP-SAVE &  HTP-DR
\\ \hline
Training error counts &    0  &  1   & 2  & NA   &  2\\ \hline
Testing error counts&     2  &   1  & 2   & NA  & 1\\ \hline
Number of genes selected &  2 &  6 & 1    & 0   & 2\\
\hline
\end{tabular}
\end{table}

\csubsection{Real data analysis}

 We consider the leukemia data
from the high-density Affymetrix oligonucleotide arrays (Golub et al, 1999).
This data set has become a benchmark in many gene expression studies. See, for example, Dettling (2004).
There are 38 training samples and 34 testing
samples, with 3571 genes in each sample. The response is $0$ or $1$
describing two subtypes of leukemia.
We first perform variable selection that is based on the training set, build a classification rule with the linear
discriminant analysis (LDA), and then apply this rule to the testing set. We compare the classification results together with the number of genes selected in Table \ref{table4}. HTP-SAVE fails to select any significant gene, suggesting that the  subtypes of leukemia
may depend on the genes through some monotone link functions. COP, SIRI, HTP-SIR and HTP-DR all lead to similar
classification performances.
While both SIRI and HTP-DR have the smallest testing error count, HTP-DR needs only 2 genes compared to 6 genes selected by SIRI.

\csection{Discussions}
For high dimensional data with unknown link functions between  predictors and  response,
it is desirable to perform variable selection in a model-free fashion. We have proposed
a versatile framework for variable selection via stepwise trace pursuit, which can be viewed as  a model-free counterpart of the classical stepwise regression.
 An important connection between sufficient dimension reduction and model-free variable selection
 is revealed in Cook (2004) via the marginal coordinate test. However, it is not applicable when $p$ is larger than $n$. Stepwise trace pursuit
 provides the missing link between sufficient dimension reduction and model-free variable selection in the high dimensional settings.
 While our discussions in this paper are based on SIR, SAVE and the directional regression, the general principle
of trace pursuit allows its extension to other sufficient dimension reduction methods  as well.

 As an important preprocessing step for ultrahigh dimensional data, variable screening is first proposed in Fan and Lv (2008) and has received much attention in the recent literature (Zhu et al., 2011; Li et al., 2012; He et al., 2013; Chang et al., 2013, Lin et al 2013). Forward trace pursuit is introduced in this paper for model-free variable screening under the sufficient dimension reduction framework.
   The screening consistency property of forward regression in linear models is established in Wang (2009), which is extended to model-free setting via SIR-based forward trace pursuit.
   %   extended in this paper for model-free variable screening via the forward trace pursuit.
%We establish
The theoretical properties of forward trace pursuit approaches that are based on other sufficient dimension reduction
methods warrant future investigation.

\newpage

\csection{Appendix: Proofs}\label{sec:proofs}

\csubsection{Proofs of Theorems in Section 2 }\label{pf sec2}

\noindent {\sc Proof of Proposition \ref{prop:trace}}. Let $\lambda_1\geq\ldots\geq\lambda_q$
be the $q$ nonzero eigenvalues of
$\M^\sir$. Denote $\bfeta_1,\ldots, \bfeta_q$
as the corresponding eigenvectors. Let $\bfbeta_i=\sig^{-1/2}\bfeta_i$ for $i=1,\ldots,q$.
Note that $\M^{\sir}=\sum_{i=1}^q \lambda_i \bfeta_i \bfeta_i^T=\sig^{1/2} \Big(\sum_{i=1}^q \lambda_i \bfbeta_i \bfbeta_i^T\Big) \sig^{1/2}$. Thus we have
  \begin{align}
 \label{eq:1}
\trace(\M^{\sir})=\trace\Big\{ \sig \Big(\sum_{i=1}^q \lambda_i \bfbeta_i \bfbeta_i^T\Big) \Big\}.
\end{align}
The  LCM assumption  (\ref{eq:lcm}) guarantees that $\bfbeta_i\in \cals_{\Y|\X}$. For $\bfbeta_i=(\beta_{i,1},\ldots,\beta_{i,p})^T$,
let $\bfbeta_{i,\cala}=\{\bfbeta_{i,j}: j\in \cala\}$ and $\bfbeta_{i,\cala^c}=\{\bfbeta_{i,j}: j\in \cala^c\}$.
 The fact that
$\Y \indep \xac|\xa$ and the definition of $\cals_{\Y|\X}$ together imply that $\bfbeta_{i,\cala^c}=\bf{0}$.
Hence (\ref{eq:1}) becomes
  \begin{align}
 \label{eq:2}
\trace(\M^{\sir})=\trace\Big\{ \sig_{\cala} \Big(\sum_{i=1}^q \lambda_i \bfbeta_{i,\cala} \bfbeta_{i,\cala}^T\Big) \Big\},
\end{align}
 where $\xa=\{\x_i: i\in \cala\}$ and $\sig_{\cala}=\var(\xa)$. By definition, we have
   \begin{align}
 \label{eq:3}
\trace (\M_{\cala}^{\sir})=\trace\{ \cov (E(\Z_{\cala}|\Y))  \}=\trace\{ \sig_{\cala}^{-1} \cov (E(\X_{\cala}|\Y))\}.
\end{align}
 % We have we $\trace(\M^{\sir})=\trace\Big\{ \sig_{\cala} \Big(\sum_{i=1}^q \lambda_i \bfbeta_{i,\cala} \bfbeta_{i,\cala}^T\Big) \Big\}$.
% Let $(\eta_1,\ldots,\eta_d)$ be the eigenvectors of $\M^{\sir}$. Let $\alpha_i=\halfinvsig\eta_i$.
%Suppose $\cala=\{1,2,\ldots,K\}$. Then $\alpha_{i,\cala^c}=0$ for $i=1,\ldots,d$. We then can write $\alpha$ as
%\begin{align*}
%\alpha=
%\begin{pmatrix}
%\alpha_{1,\cala} &\ldots & \alpha_{d,\cala} \\
%0 & \ldots & 0
%\end{pmatrix}
%\end{align*}
%Then we could see that
%\begin{align*}
%\trace(\M^{\sir})=&\trace\Big(\sum_{i=1}^q \lambda_i \bfeta_i \bfeta_i^T\Big) = \trace\Big\{\sig^{1/2} \Big(\sum_{i=1}^q \lambda_i \bfbeta_i \bfbeta_i^T\Big) \sig^{1/2}\Big\}\\
%=& \trace\Big\{\sig \Big(\sum_{i=1}^d \lambda_i \alpha_i \alpha_i^T\Big)\Big\} = \trace\Big\{\sig_{\cala} \Big(\sum_{i=1}^d \lambda_{i} \alpha_{i,\cala} \alpha_{i,\cala}^T\Big) \Big\}
%\end{align*}
Assume $\cala=\{1,2,\ldots,K\}$ without loss of generality.
Note that
\begin{align*}
&\cov (E(\X|\Y)) =  \sig^{1/2} \M^{\sir} \sig^{1/2} = \sig \Big(\sum_{i=1}^q \lambda_i \bfbeta_i \bfbeta_i^T \Big)\sig\\
&=
\begin{pmatrix}
\sig_{\cala} & \sig_{\cala,\cala^c}\\
\sig_{\cala^c,\cala} & \sig_{\cala^c}
\end{pmatrix}
\begin{pmatrix}
\sum_{i=1}^q \lambda_i \bfbeta_{i,\cala} \bfbeta_{i,\cala}^T & \bf{0}\\
\bf{0} & \bf{0}
\end{pmatrix}
\begin{pmatrix}
\sig_{\cala} & \sig_{\cala,\cala^c}\\
\sig_{\cala^c,\cala} & \sig_{\cala^c}
\end{pmatrix},
\end{align*}
the upper left block of which implies that $ \cov (E(\X_{\cala}|\Y))=\sig_{\cala}\left(\sum_{i=1}^q \lambda_i \bfbeta_{i,\cala} \bfbeta_{i,\cala}^T \right)  \sig_{\cala}$. Plug it into (\ref{eq:3}), and we get $\trace(\M_{\cala}^{\sir})=\trace\Big\{ \sig_{\cala} \Big(\sum_{i=1}^q \lambda_i \bfbeta_{i,\cala} \bfbeta_{i,\cala}^T\Big) \Big\}$. Together with (\ref{eq:2}), we have $\trace(\M_{\cala}^{\sir})=\trace(\M^{\sir})$.
For $\calf$ satisfying $\cala\subseteq \calf$, the proof is similar and omitted.
 \eop

\noindent {\sc Proof of Theorem \ref{prop:sir}}. Condition (\ref{eq:subset lcm}) implies that for $\calf\subset \cali$ and $j\in \calf^c$, we have $E(\x_j | \xf)=E^T(\x_j\xf)\sig_{\calf}^{-1}\xf$. Recall that
 $|\calf|$ denotes the cardinality of $\calf$. For the first part, define $(|\calf|+1)\times(|\calf|+1)$ dimensional matrices $\A$ and $\C$
 as
  \begin{align}
 \label{eq:transformation a}
\A=
\begin{pmatrix}
\I_{|\calf|} & \0 \\
-E^T(\x_j\xf)\sig_{\calf}^{-1} & 1
\end{pmatrix}
\mbox{ and }
\C=
\begin{pmatrix}
\sig_{\calf} & \0 \\
\0 & \sigma_{j|\mathcal{F}}^2
\end{pmatrix}.
%\end{pmatrix}.
 \end{align}
 Recall that $\U_h=E(\X|\Y \in J_h)$.
It is easy to check that
% $\X_{(\calf,j)}^T\A^T=
% (\xf^T,\x_j) \A^T=(\xf^T,\x_{j|\mathcal{F}})$.
  \begin{align}
  \label{eq:transformation a1}
\A\X_{\calf\cup j}=
\begin{pmatrix}
\xf \\
\x_{j|\mathcal{F}}
\end{pmatrix}
\mbox{ and }
\A \U_{\calf\cup j,h}=
\begin{pmatrix}
\U_{\calf,h} \\
E(\x_{j|\mathcal{F}}|\Y \in J_h)
\end{pmatrix}.
 \end{align}
 Because
 $E(\xf \x_{j|\mathcal{F}})=E\{\xf E(\x_{j|\mathcal{F}}|\xf)\}=0$, we have
 $\var(\A\X_{\calf\cup j})=\A\sig_{\calf\cup j}\A^T=\C$. Hence $\sig_{\calf\cup j}^{-1}=\A^T \C^{-1}\A$.
%   \begin{align}
%  \label{eq:transformation a3}
%\sig_{\calf\cup j}^{-1}=\A^T \C^{-1}\A.
% \end{align}
%  From (\ref{eq:kernel sir}) we have
Together with $\trace(\M^\sir_{\calf\cup j})=\trace\left\{\sig_{\calf\cup j}^{-1}\left(\sum_{h=1}^H p_h
\U_{\calf\cup j,h}\U_{\calf\cup j,h}^T\right) \right\}$, we get
% Plugging in (\ref{eq:transformation a3}) and we get
  \begin{align}
  \label{eq:transformation a2}
  \trace(\M^\sir_{\calf\cup j})=\trace\left(\C^{-1}\left\{\sum_{h=1}^H p_h
(\A \U_{\calf\cup j,h}) (\A \U_{\calf\cup j,h})^T\right\} \right).
 \end{align}
Plug (\ref{eq:transformation a}) and (\ref{eq:transformation a1}) into (\ref{eq:transformation a2}), and we get
  \begin{align*}
%  \label{eq:transformation a2}
  \trace(\M^\sir_{\calf\cup j})=\trace\left\{\sig_{\calf}^{-1}\left(\sum_{h=1}^H p_h
\U_{\calf,h}\U_{\calf,h}^T\right) \right\}+\sum_{h=1}^H p_h \sigma_{j|\mathcal{F}}^{-2}E^2(\x_{j|\mathcal{F}}|\Y \in J_h).
 \end{align*}
It follows immediately that $\trace(\M^\sir_{\calf\cup j})-\trace(\M^\sir_\calf)=\sum_{h=1}^H p_h
\gamma_{j|\mathcal{F},h}^2$.

For the second part, note that $\Y \indep \xac|\xa$, $\cala\subseteq\calf$ and $j\in \calf^c$ together imply that $\Y\indep \x_j|\xf$. Thus we have $E(\x_j|\Y,\xf)=E(\x_j|\xf)$. It follows that
\begin{align}
\label{eq:sir key}
E(\x_j| \Y)=E\{ E(\x_j|\Y,\xf)|\Y\}=E\{E(\x_j|\xf)|\Y\}.
\end{align}
 As a result $E(\x_{j|\mathcal{F}}|\Y)=E[\{\x_j-E(\x_j|\xf)\}|\Y]=0$ and
 $\gamma_{j|\mathcal{F},h}=0$. Hence $\trace(\M^\sir_{\calf\cup j})-\trace(\M^\sir_\calf)=0$.
\eop

\noindent {\sc Proof of Theorem \ref{prop:save}}.
Define $\A$ and $\C$ as in (\ref{eq:transformation a}).
Denote $\B_{\calf,h}=\sig_{\calf}-\V_{\calf, h}+\U_{\calf, h}\U_{\calf, h}^T$.
 It follows that
\begin{align*}
\trace(\M^{\save}_{\calf\cup j})=\sum_{h=1}^H p_h\trace\left(\sig_{\calf\cup j}^{-1} \B_{\calf\cup j,h}\sig_{\calf\cup j}^{-1} \B_{\calf\cup j,h} \right).
\end{align*}
By noticing $\var(\X_{\calf\cup j})=\sig_{\calf\cup j}$ and $\var(\A\X_{\calf\cup j})=\C$, we have
 $\sig_{\calf\cup j}^{-1}=\A^T \C^{-1}\A$. Let $\D_h=\C^{-1/2}\A \B_{\calf\cup j,h} \A^T \C^{-1/2}$. Then
 \begin{align*}
\trace(\M^{\save}_{\calf\cup j})=\sum_{h=1}^H p_h\trace\left(\C^{-1}\A \B_{\calf\cup j,h} \A^T \C^{-1}\A \B_{\calf\cup j,h} \A^T \right)=\sum_{h=1}^H p_h\trace(\D_h\D_h).
\end{align*}
To calculate $\A \B_{\calf\cup j,h}\A^T=\A\left(\sig_{\calf\cup j}-\V_{\calf\cup j, h}+\U_{\calf\cup j, h}\U_{\calf\cup j, h}^T\right)\A^T$, note that
\begin{align*}
&\A\sig_{\calf\cup j}\A^T=
%\C=
\begin{pmatrix}
\sig_{\calf} & \0 \\
\0 & \sigma_{j|\mathcal{F}}^2
\end{pmatrix},
\A\V_{\calf\cup j, h}\A^T=\begin{pmatrix}
\V_{\calf, h} & E(\xf\x_{j|\mathcal{F}}|\Y
\in J_h) \\
E^T(\xf\x_{j|\mathcal{F}}|\Y
\in J_h) & E(\x^2_{j|\mathcal{F}}|\Y \in J_h)
\end{pmatrix},\\
&\mbox{ and }\A\U_{\calf\cup j, h}\U_{\calf\cup j, h}^T\A^T=\begin{pmatrix}
\U_{\calf, h}\U_{\calf, h}^T & \U_{\calf,h}E(\x_{j|\mathcal{F}}|\Y \in J_h) \\
  \U^T_{\calf,h}E(\x_{j|\mathcal{F}}|\Y \in J_h)  & E^2(\x_{j|\mathcal{F}}|\Y \in J_h)
\end{pmatrix}.
\end{align*}
Let $\bfpsi_{j|\mathcal{F},h}=\U_{\calf,h}E(\x_{j|\mathcal{F}}|\Y \in J_h)-
E(\xf\x_{j|\mathcal{F}}|\Y
\in J_h)$  and $D_{22}=\sigma_{j|\mathcal{F}}^{2}$ $-$ $E(\x^2_{j|\mathcal{F}}|\Y \in J_h)+
E^2(\x_{j|\mathcal{F}}|\Y \in J_h)$.
Then
\begin{align*}
\A \B_{\calf\cup j,h}\A^T=
 \begin{pmatrix}
\B_{\calf,h} &\bfpsi_{j|\mathcal{F},h}\\
\bfpsi_{j|\mathcal{F},h}^T &  D_{22}
\end{pmatrix}
%\mbox{ and }
\end{align*}
It follows that $\D_h=\C^{-1/2}\A \B_{\calf\cup j,h} \A^T \C^{-1/2}$ becomes
\begin{align*}
\D_h=
\begin{pmatrix}
\sig_{\calf}^{-1/2}\B_{\calf,h}\sig_{\calf}^{-1/2} &\bfphi_{j|\mathcal{F},h}\\
\bfphi_{j|\mathcal{F},h}^T &  1-\zeta_{j|\mathcal{F},h}+\gamma_{j|\mathcal{F},h}^2
\end{pmatrix}.
\end{align*}
The conclusion of the first part is then obvious.

For the second part, we now show that $\bfphi_{j|\mathcal{F},h}=\0$, $\gamma_{j|\mathcal{F},h}=0$,
and $\zeta_{j|\mathcal{F},h}=1$. Note that $\gamma_{j|\mathcal{F},h}=0$ in the proof of Theorem \ref{prop:sir}.
Similar to (\ref{eq:sir key}) where we have shown
$E(\x_j| \Y)=E\{E(\x_j|\xf)|\Y\}$, it can be shown that
$E(\x_j\xf| \Y)=E\{\xf E(\x_j|\xf)|\Y\}$. Hence $E(\x_{j|\mathcal{F}}|\Y)=0$ and $E(\xf\x_{j|\mathcal{F}}|\Y)=\0$.
It follows that $\cov(\xf,\x_{j|\mathcal{F}}|\Y)=\0$, and $\bfphi_{j|\mathcal{F},h}=-\sig_{\calf}^{-1/2}\cov(\xf,\x_{j|\mathcal{F}}|\Y\in J_h)=\0$.
It remains to show that $\zeta_{j|\mathcal{F},h}=E(\gamma_{j|\mathcal{F}}^2|\Y$ $\in J_h)=1$, or $E(\x_{j|\calf}^2|\Y)=\var(\x_{j|\calf})$.
Because  $\x_{j|\mathcal{F}}=\x_j-E(\x_j|\xf)$ and
$\cov\{\x_j,E(\x_j|\xf)\}=E\{\x_jE(\x_j|\xf)\}=\var\{E(\x_j|\xf)\}$,
we have
\begin{align}
\label{save1}
\var(\x_{j|\calf})=\var(\x_j)-\var\{E(\x_j|\xf)\}=E\{\var(\x_j|\xf)\}.
\end{align}
%where the first equality follows from
%$$\cov\{\x_j,E(\x_j|\xf)\}=E\{\x_jE(\x_j|\xf)\}=\var\{E(\x_j|\xf)\}.$$
The
subset CCV condition (\ref{eq:subset ccv}) implies that
$E\{(\x_j-E(\x_j|\xf))^2  |\Y,\xf\}=E\{(\x_j-E(\x_j|\xf))^2  |\xf\}=\var(\x_j|\xf)$.
Thus we have
 \begin{align}
\label{save2}
E(\x_{j|\calf}^2|\Y)=E\{(\x_j-E(\x_j|\xf))^2  |\Y,\xf\}=E\{\var(\x_j|\xf)|\Y\}.
\end{align}
Compare
(\ref{save1}) with
(\ref{save2}) and we get the desired result.\eop

\noindent {\sc Proof of Theorem \ref{prop:dr}}. Let $\M^{\dr 1}_{\mathcal{F}\cup j}=\sum_{h=1}^H p_h
\left(\I_{|\calf|+1}-\sig^{-1/2}_{\mathcal{F}\cup j}
\V_{\mathcal{F}\cup j,h}
\sig^{-1/2}_{\mathcal{F}\cup j}\right)^2$, $\M^{\dr 2}_{\mathcal{F}\cup j}=\sum_{h=1}^H p_h
\sig^{-1/2}_{\mathcal{F}\cup j}\U_{\mathcal{F}\cup j,h} \U_{\mathcal{F}\cup j,h}^T \sig^{-1/2}_{\mathcal{F}\cup j}$,
and $m^{\dr 3}_{\mathcal{F}\cup j}=\sum_{h=1}^H p_h \U_{\mathcal{F}\cup j,h}^T \sig^{-1}_{\mathcal{F}\cup j} \U_{\mathcal{F}\cup j,h}$.
Then $\M^\dr_{\mathcal{F}\cup j}/2$ can be written as
\begin{align}
\label{dr proof}
\M^{\dr 1}_{\mathcal{F}\cup j}+\left(\M^{\dr 2}_{\mathcal{F}\cup j}\right)^2+ m^{\dr 3}_{\mathcal{F}\cup j}\M^{\dr 2}_{\mathcal{F}\cup j}.
\end{align}
%With
% $\M^{\dr 1}_{\mathcal{F}\cup j}=\sum_{h=1}^H p_h
%\left(\I_{|\calf|+1}-\sig^{-1/2}_{\mathcal{F}\cup j}
%\V_{\mathcal{F}\cup j,h}
%\sig^{-1/2}_{\mathcal{F}\cup j}\right)^2$,
The first term in (\ref{dr proof}) can be shown to satisfy
$$\trace(\M^{\dr 1}_{\mathcal{F}\cup j})=\trace(\M^{\dr 1}_{\mathcal{F}})+\sum_{h=1}^H
p_h\left\{(1-\zeta_{j|\mathcal{F},h})^2+
2\bfnu_{j|\mathcal{F},h}^T\bfnu_{j|\mathcal{F},h}\right\}.$$
%With $\M^{\dr 2}_{\mathcal{F}\cup j}=\sum_{h=1}^H p_h
%\sig^{-1/2}_{\mathcal{F}\cup j}\U_{\mathcal{F}\cup j,h} \U_{\mathcal{F}\cup j,h}^T \sig^{-1/2}_{\mathcal{F}\cup j}$,
The second term in (\ref{dr proof}) can be shown to satisfy
$$\trace\left\{(\M^{\dr 2}_{\mathcal{F}\cup j})^2\right\}=\trace\left\{(\M^{\dr 2}_{\mathcal{F}})^2\right\}+2\bfiota_{j|\mathcal{F}}^T \bfiota_{j|\mathcal{F}}
%\left(\sum_{h=1}^H p_h \iota_{j|\mathcal{F},h}^T\right)\left(\sum_{h=1}^H p_h \iota_{j|\mathcal{F},h}\right)
%+\left(\sum_{h=1}^H p_h \gamma^2_{j|\mathcal{F},h}\right)^2
+\varrho_{j|\mathcal{F}}^2.$$
%With $m^{\dr 3}_{\mathcal{F}\cup j}=\sum_{h=1}^H p_h \U_{\mathcal{F}\cup j,h}^T \sig^{-1}_{\mathcal{F}\cup j} \U_{\mathcal{F}\cup j,h}$,
The last term in (\ref{dr proof}) can be shown to satisfy
$$\trace\left(m^{\dr 3}_{\mathcal{F}\cup j}\M^{\dr 2}_{\mathcal{F}\cup j}\right)=\trace\left(m^{\dr 3}_{\mathcal{F}}\M^{\dr 2}_{\mathcal{F}}\right)+2\kappa_\calf
\varrho_{j|\mathcal{F}}+\varrho_{j|\mathcal{F}}^2.$$
Together we get the first part of Theorem \ref{prop:dr}.
%\begin{align*}
%&\M^{\dr 1}_{\mathcal{F}\cup j}=\sum_{h=1}^H p_h
%\left(\I_{|\calf|+1}-\sig^{-1/2}_{\mathcal{F}\cup j}
%\V_{\mathcal{F}\cup j,h}
%\sig^{-1/2}_{\mathcal{F}\cup j}\right)^2,\\
%&\M^{\dr 2}_{\mathcal{F}\cup j}=\sum_{h=1}^H p_h
%\sig^{-1/2}_{\mathcal{F}\cup j}\U_{\mathcal{F}\cup j,h} \U_{\mathcal{F}\cup j,h}^T \sig^{-1/2}_{\mathcal{F}\cup j},\\
%&m^{\dr 3}_{\mathcal{F}\cup j}=\sum_{h=1}^H p_h \U_{\mathcal{F}\cup j,h}^T \sig^{-1}_{\mathcal{F}\cup j} \U_{\mathcal{F}\cup j,h} .
%\end{align*}
%Similar to the proofs of Theorems \ref{prop:sir} and \ref{prop:save}, we can show that
%\begin{align*}
%&\trace(\M^{\dr 1}_{\mathcal{F}\cup j})=\trace(\M^{\dr 1}_{\mathcal{F}})+\sum_{h=1}^H
%p_h\left((1-\zeta_{j|\mathcal{F},h})^2+
%2\bfnu_{j|\mathcal{F},h}^T\bfnu_{j|\mathcal{F},h}\right),\\
%&\trace\left\{(\M^{\dr 2}_{\mathcal{F}\cup j})^2\right\}=\trace\left\{(\M^{\dr 2}_{\mathcal{F}})^2\right\}+2\bfiota_{j|\mathcal{F}}^T \bfiota_{j|\mathcal{F}}
%%\left(\sum_{h=1}^H p_h \iota_{j|\mathcal{F},h}^T\right)\left(\sum_{h=1}^H p_h \iota_{j|\mathcal{F},h}\right)
%%+\left(\sum_{h=1}^H p_h \gamma^2_{j|\mathcal{F},h}\right)^2
%+\varrho_{j|\mathcal{F}}^2,\\
%&\trace\left(m^{\dr 3}_{\mathcal{F}\cup j}\M^{\dr 2}_{\mathcal{F}\cup j}\right)=\trace\left(m^{\dr 3}_{\mathcal{F}}\M^{\dr 2}_{\mathcal{F}}\right)+2\kappa_\calf
%\varrho_{j|\mathcal{F}}+\varrho_{j|\mathcal{F}}^2.
%%\left(\sum_{h=1}^H p_h \gamma^2_{j|\mathcal{F},h}\right)+\left(\sum_{h=1}^H p_h \gamma^2_{j|\mathcal{F},h}\right)^2.
%\end{align*}
For the second part, we have seen in the proof of Theorem \ref{prop:save}
that $\gamma_{j|\mathcal{F},h}=0$,
and $\zeta_{j|\mathcal{F},h}=1$ given that $\cala\subseteq\calf$. It is easy to see that  $\bfnu_{j|\mathcal{F},h}=\0$,
$\bfiota_{j|\mathcal{F}}=\0$ and $\varrho_{j|\mathcal{F}}=0$. The conclusion is then obvious.
\eop

\csubsection{ Proofs of Theorems in Section 3}\label{pf sec3}

We use Frechet derivative representation to derive the asymptotic distributions of $T^\sir_{j|\mathcal{F}}$,
$T^\save_{j|\mathcal{F}}$ and $T^\dr_{j|\mathcal{F}}$.
Let $F$ be the joint distribution of $(\X,Y)$ and $F_n$ be the empirical distribution based on the i.i.d. sample
$(\X_1,Y_1),\ldots,(\X_n,Y_n)$. Let $G$ be a real or matrix valued functional. Then $G(F_n)$ has the following asymptotic expansion under regularity conditions,
\begin{align}
\label{frechet}
G(F_n)=G(F)+E_n \{G^*(F)\}+O_p(n^{-1}),
\end{align}
where $G(F)$ is nonrandom, and $E_n \{G^*(F)\}=O_p(n^{-1/2})$ as $G^*(F)$ satisfies $E\{G^*(F)\}=0$.
Please refer to Fernholz (1983) for more details about Frechet derivative and the regularity conditions.

Recall the following definitions involved in  Theorem \ref{theorem:sir}: $R_h=I(\Y\in J_h)$, $p_h=E(I(\Y\in J_h))$,
%$\U_h=E(\X|\Y \in J_h)$,
$\sig_{\calf}=\var(\xf)$, $\U_{\calf,h}=E(\xf|\Y \in J_h)$,
$\x_{j|\mathcal{F}}=\x_j-E(\x_j|\xf)$,
$\sigma_{j|\mathcal{F}}^2=\var(\x_{j|\mathcal{F}})$,
$\gamma_{j|\mathcal{F}}=\x_{j|\mathcal{F}}/\sigma_{j|\mathcal{F}}$,
and
$\gamma_{j|\mathcal{F},h}=E(\gamma_{j|\mathcal{F}}|\Y \in J_h)$. Denote $u_{j,h}=E(x_j|\Y \in J_h)$ and $\bftheta_{j|\calf}=\sig_{\calf}^{-1}E(\x_j\xf)$. Then for $\calf\subset \cali$ and $j\in \calf^c$,
condition (\ref{eq:subset lcm}) implies $E(\x_j | \xf)=\bftheta_{j|\calf}^T \xf$.

\begin{lemma}\label{lemma:one}
Assume the conditions of Theorem \ref{theorem:sir} are satisfied. Under $H_0:\Y \indep \x_j| \xf, j\in \calf^c$, the
expansions of
%$\hat p_h$,
$\hat \sig_{\calf}$, $\hat \sig_{\calf}^{-1}$, $\hat \U_{\calf,h}$, $\hat \bftheta_{j|\calf}$,
  %$\hat \sigma^2_{j|\mathcal{F}}$
 and $\hat \gamma_{j|\mathcal{F},h}$
  have the form (\ref{frechet}), where we
  replace $G(F)$ with
  %$p_h$,
  $\sig_{\calf}$, $\sig_{\calf}^{-1}$, $\U_{\calf,h}$, $\bftheta_{j|\calf}$, or
  %$\sigma^2_{j|\mathcal{F}}$ or
  $\gamma_{j|\mathcal{F},h}$, and we replace $G^*(F)$ with
  %$p_h^*$, $\sig_{\calf}^*$, $(\sig_{\calf}^{-1})^*$, $\U_{\calf,h}^*$, $\bftheta_{j|\calf}^*$,
  %$(\sigma^2_{j|\mathcal{F}})^*$,
  %$\gamma_{j|\mathcal{F},h}^*$.
%$p_h^*=R_h-p_h$,
 $\sig_{\calf}^*=\xf\xf^T-\sig_{\calf}$, $(\sig_{\calf}^{-1})^*=-\sig_{\calf}^{-1}\sig_{\calf}^*\sig_{\calf}^{-1}$,
$\U_{\calf,h}^*=(\xf-\U_{\calf,h}) R_h/p_h-\xf$, $\bftheta^*_{j|\calf}=\sig_{\calf}^{-1} \{\x_j\xf-E(\x_j\xf)\}
+ \left(\sig_{\calf}^{-1}\right)^*E(\x_j\xf)$, or
%$(\sigma^2_{j|\mathcal{F}})^*=
%(-\bftheta^T_{j|\calf},1)\sig^*_{\calf\cup j}(-\bftheta_{j|\calf}^T,1)^T+2(-(\bftheta^*_{j|\calf})^T,0)
%\sig_{\calf\cup j}(-\bftheta_{j|\calf}^T,1)^T$, or
$\gamma_{j|\mathcal{F},h}^*=\{u^*_{j,h}-(\bftheta_{j|\mathcal{F}}^*)^T \U_{\calf,h}
-\bftheta_{j|\calf}^T \U^*_{\calf,h}\}/{\sigma_{j|\mathcal{F}}}$.
\end{lemma}

\noindent {\sc Proof of Lemma \ref{lemma:one}.} The
expansions of
%$\hat p_h$,
$\hat \sig_{\calf}$, $\hat \sig_{\calf}^{-1}$, $\hat \U_{\calf,h}$ and $\bftheta_{j|\calf}$
are similar to those in Lemma 1 in Li and Wang (2007), and thus omitted. With condition (\ref{eq:subset lcm}),
we have $\x_{j|\mathcal{F}}=\x_j-E(\x_j|\xf)=\x_j-\bftheta_{j|\calf}^T \xf$.
%Thus $\sigma_{j|\mathcal{F}}^2=\var(\x_{j|\mathcal{F}})$ can be rewritten as $\sigma_{j|\calf}^2=(-\bftheta^T_{j|\calf},1)\sig_{\calf\cup j}(-\bftheta_{j|\calf}^T,1)^T$, and the expansion of $\hat \sigma^2_{j|\mathcal{F}}$ is obvious.
For the expansion of $\hat\gamma_{j|\mathcal{F},h}$, notice that
\begin{align*}
%\label{lemma eq 1}
\sigma_{j|\mathcal{F}} \gamma_{j|\mathcal{F},h}=
E(\x_j-\bftheta_{j|\calf}^T\xf|\Y \in J_h)=u_{j,h}-\bftheta_{j|\calf}^T \U_{\calf,h}.
\end{align*}
We have seen in the proof of Theorem \ref{prop:sir} that $\gamma_{j|\mathcal{F},h}=0$ with  $\Y \indep \x_j| \xf$ and condition (\ref{eq:subset lcm}).
 Taking
Frechet derivative on both sides of the listed equation above, we get $\sigma_{j|\mathcal{F}}
\gamma_{j|\mathcal{F},h}^*= u_{j,h}^*-(\bftheta_{j|\mathcal{F}}^*)^T \U_{\calf,h}
-\bftheta_{j|\calf}^T \U^*_{\calf,h}$. \eop

Note that
$u_{j,h}$ is a special case of $\U_{\calf,h}$ with $\calf$ replaced by $j$. Thus expansion of $\hat u_{j,h}$
follows the same form of $\hat \U_{\calf,h}$ and is omitted.

\noindent {\sc Proof of Theorem \ref{theorem:sir}.}
Let $\hat\bfell^\sir_{j|\mathcal{F}}=(\hat p_1^{1/2}
\hat \gamma_{j|\mathcal{F},1},\ldots,\hat p_H^{1/2}\hat \gamma_{j|\mathcal{F},H})^T$. Then
$T^\sir_{j|\mathcal{F}}=n\sum_{h=1}^H \hat p_h
\hat\gamma_{j|\mathcal{F},h}^2$ can be written as
$n (\hat\bfell^\sir_{j|\mathcal{F}})^T \hat\bfell^\sir_{j|\mathcal{F}}$. Because  $\gamma_{j|\mathcal{F},h}=0$ with  $\Y \indep \x_j| \xf$ and condition (\ref{eq:subset lcm}), $\hat \bfell^\sir_{j|\mathcal{F}}$ has expansion
$$\hat \bfell^\sir_{j|\mathcal{F}}=\bfell^\sir_{j|\mathcal{F}}+E_n\{(\bfell^\sir_{j|\mathcal{F}})^*\}+o_P(n^{-1/2}),$$
where $(\bfell^\sir_{j|\mathcal{F}})^*=(p_1^{1/2}
\gamma_{j|\mathcal{F},1}^*,\ldots,p_H^{1/2}\gamma_{j|\mathcal{F},H}^*)^T$, and
$\gamma_{j|\mathcal{F},h}^*$ is provided in Lemma \ref{lemma:one}.
 Define
${\bf
\Omega}^\sir_{j|\mathcal{F}}=E((\bfell^\sir_{j|\mathcal{F}})^*\{(\bfell^\sir_{j|\mathcal{F}})^*\}^T)$,
and the result of Theorem \ref{theorem:sir} follows directly.\eop

Recall the following definitions involved in  Theorem \ref{theorem:save}:
 $\bfnu_{j|\mathcal{F},h}=\sig^{-1/2}_{\mathcal{F}}E(\xf\gamma_{j|\mathcal{F}}|\Y
\in J_h)$, $\zeta_{j|\mathcal{F},h}=E(\gamma_{j|\mathcal{F}}^2|\Y\in J_h)$, $\bfphi_{j|\mathcal{F},h}=\sig_{\calf}^{-1/2}\left\{ \U_{\calf, h}\gamma_{j|\mathcal{F},h}-E(\xf\gamma_{j|\mathcal{F}}|\Y
\in J_h)\right\}$, and $\V_{\calf, h}=E(\xf$ $\xf^T|\Y \in
J_h)$.

\begin{lemma}\label{lemma:two}
Assume the conditions of Theorem \ref{theorem:save} are satisfied. Under $H_0:\Y \indep \x_j| \xf$, $j\in \calf^c$,  the
expansions of $\hat \V_{\calf,h}$, $\hat\bfnu_{j|\mathcal{F},h}$, $\hat\bfphi_{j|\mathcal{F},h}$, $\hat\zeta_{j|\mathcal{F},h}$
  have the form (\ref{frechet}), where we
  replace $G(F)$ with $\V_{\calf,h}$, $\bfnu_{j|\mathcal{F},h}$, $\bfphi_{j|\mathcal{F},h}$, or $\zeta_{j|\mathcal{F},h}$, and we replace $G^*(F)$ with
  $\V_{\calf,h}^*=(\xf\xf^T
R_h-\V_{\calf,h})/p_h-\xf(\U_{\calf,h}^*)^T-(\U_{\calf,h}^*)\xf^T$,  $\bfnu_{j|\calf,h}^*=\sigf^{-1/2}(\x_j\xf-E(\x_j\xf)-\V^*_{\calf}\bftheta_{j|\calf}-\V_{\calf}\bftheta_{j|\calf}^*)/\sigma_{j|\calf}$, $\bfphi_{j|\calf,h}^*=\sigf^{-1/2}\U_{\calf,h}\gamma_{j|\calf,h}^*-\bfnu_{j|\calf,h}^*$, or
 $\zeta_{j|\calf,h}^*=\{(-\bftheta^T_{j|\calf},1)$ $\V^*_{\calf\cup j,h}(-\bftheta_{j|\calf}^T,1)^T+2(-(\bftheta^*_{j|\calf})^T,1)\V_{\calf\cup j,h}(-\bftheta_{j|\calf}^T,1)^T-(\sigma^2_{j|\mathcal{F}})^*\}/\sigma^2_{j|\mathcal{F}}$.
\end{lemma}

\noindent {\sc Proof of Lemma \ref{lemma:two}.} The
expansion of $\hat \V_{\calf,h}$ is parallel to the expansion of $\hat \V_{h}$ in Lemma 1 of Li and Wang (2007).
 The expansion of $\hat\bfnu_{j|\mathcal{F},h}$ uses the same technique as the expansion of $\hat\gamma_{j|\mathcal{F},h}$ in Lemma \ref{lemma:one}. The expansion of $\hat\bfphi_{j|\mathcal{F},h}$ is obvious by noticing that $\gamma_{j|\mathcal{F},h}=0$ with  $\Y \indep \x_j| \xf$ and condition (\ref{eq:subset lcm}).
For the expansion of $\hat\zeta_{j|\mathcal{F},h}$, notice that
\begin{align*}
%\label{lemma eq 2}
\sigma^2_{j|\mathcal{F}}\zeta_{j|\calf,h}
= E\big((\x_j-\bftheta^T_{j|\calf}\X_{\calf})^2|\Y\in J_h\big)=(-\bftheta^T_{j|\calf},1)\V_{\calf\cup j,h}(-\bftheta_{j|\calf}^T,1)^T.
\end{align*}
We have seen in the proof of Theorem \ref{prop:save} that $\zeta_{j|\calf,h}=1$ with  $\Y \indep \x_j| \xf$,  conditions (\ref{eq:subset lcm}) and (\ref{eq:subset ccv}). Taking
Frechet derivative on both sides of the listed equation above, we get
the desired result. \eop

\noindent {\sc Proof of Theorem \ref{theorem:save}.}
Let $(\hat\bfell^{\save}_{j|\mathcal{F}})^T$ $=$ $\{(\hat\bfell^{\save 1}_{j|\mathcal{F}})^T, (\hat\bfell^{\save 2}_{j|\mathcal{F}})^T\}$ with $\hat\bfell^{\save 1}_{j|\mathcal{F}}=\{\hat p_1^{1/2}
(1-\hat\zeta_{j|\mathcal{F},1}+\hat\gamma_{j|\mathcal{F},1}^2),\ldots,
\hat p_H^{1/2}(1-\hat\zeta_{j|\mathcal{F},H}+\hat\gamma_{j|\mathcal{F},H}^2)\}^T$ and
 $\hat\bfell^{\save 2}_{j|\mathcal{F}}=(\sqrt{2}\hat p_1^{1/2}\hat\bfphi_{j|\mathcal{F},1}^T,\ldots,$ $\sqrt{2}\hat p_H^{1/2}\hat\bfphi_{j|\mathcal{F},H}^T )^T$.
Then
$T^\save_{j|\mathcal{F}}
 =n (\hat\bfell^{\save}_{j|\mathcal{F}})^T \hat\bfell^{\save}_{j|\mathcal{F}}$.
 Let
   $(\bfell^{\save 1}_{j|\mathcal{F}})^*=(p_1^{1/2}
\zeta_{j|\mathcal{F},1}^*,\ldots,$ $p_H^{1/2}\zeta_{j|\mathcal{F},H}^*)^T$,
 $(\bfell^{\save 2}_{j|\mathcal{F}})^*=\{\sqrt{2}p_1^{1/2}(\bfphi_{j|\mathcal{F},1}^*)^T$, $\ldots$, $\sqrt{2}p_H^{1/2}(\bfphi_{j|\mathcal{F},H}^*)^T \}^T$, and $\{(\bfell^{\save}_{j|\mathcal{F}})^*\}^T=(\{(\bfell^{\save 1}_{j|\mathcal{F}})^*\}^T,\{(\bfell^{\save 2}_{j|\mathcal{F}})^*\}^T )$. Here $\bfphi_{j|\mathcal{F},h}^*$ and $\zeta_{j|\mathcal{F},h}^*$
  are
  provided in Lemma \ref{lemma:two}. With $\Y \indep \x_j| \xf$,  conditions (\ref{eq:subset lcm}) and (\ref{eq:subset ccv}), we have
  $\gamma_{j|\mathcal{F},h}=0$ and $\zeta_{j|\mathcal{F},h}=1$. It follows that
$$\hat \bfell^{\save}_{j|\mathcal{F}}=\bfell^{\save}_{j|\mathcal{F}}+E_n\{(\bfell^{\save}_{j|\mathcal{F}})^*\}+o_P(n^{-1/2}).$$
Let ${\bf
\Omega}^{\textup{SAVE}}_{j|\mathcal{F}}=E((\bfell^{\save}_{j|\mathcal{F}})^*\{(\bfell^{\save}_{j|\mathcal{F}})^*\}^T )$.
The result of Theorem \ref{theorem:save} follows directly.\eop

Recall the following definitions involved in  Theorem \ref{theorem:dr}: $\bfiota_{j|\mathcal{F},h}=\sig^{-1/2}_{\mathcal{F}}$ $\U_{\calf,h}\gamma_{j|\mathcal{F},h}$,
$\bfiota_{j|\mathcal{F}}=\sum_{h=1}^H p_h
\bfiota_{j|\mathcal{F},h}$, $\varrho_{j|\mathcal{F}}=\sum_{h=1}^H p_h
\gamma^2_{j|\mathcal{F},h}$,
and $\kappa_{\mathcal{F}}=\sum_{h=1}^H p_h$ $\U_{\calf,h}^T
\sig^{-1}_{\mathcal{F}} \U_{\calf,h}$.   Proof of the following lemma is obvious and omitted.

\begin{lemma}\label{lemma:three}
Assume the conditions of Theorem \ref{theorem:dr} are satisfied. Under $H_0:\Y \indep \x_j| \xf, j\in \calf^c$,  the
expansion of $\hat \bfiota_{j|\mathcal{F},h}$
  has the form (\ref{frechet}), where we
  replace $G(F)$ with $\bfiota_{j|\mathcal{F},h}$,  and we replace $G^*(F)$ with
  $\bfiota_{j|\mathcal{F},h}^*=\sigf^{-1/2}\U_{\calf,h}\gamma_{j|\mathcal{F},h}^*$.
  \end{lemma}

\noindent {\sc Proof of Theorem \ref{theorem:dr}.}
 Let $\hat\bfell^{\dr 1}_{j|\mathcal{F}}=\{\sqrt{2}\hat p_1^{1/2}
(1-\hat\zeta_{j|\mathcal{F},1}),\ldots,\sqrt{2}\hat p_H^{1/2}(1-\hat\zeta_{j|\mathcal{F},H})\}^T$,
 $\hat\bfell^{\dr 2}_{j|\mathcal{F}}=\{2\hat p_1^{1/2}\hat\bfnu_{j|\mathcal{F},1}^T, \ldots, 2\hat p_H^{1/2}\hat\bfnu_{j|\mathcal{F},H}^T \}^T$,
 $\hat\ell^{\dr 3}_{j|\mathcal{F}}=2\sum_{h=1}^H \hat p_h
\hat \gamma^2_{j|\mathcal{F},h}$,
 $\hat\bfell^{\dr 4}_{j|\mathcal{F}}=2\sum_{h=1}^H\hat p_h
\hat\bfiota_{j|\mathcal{F},h}$, and
$\hat\bfell^{\dr 5}_{j|\mathcal{F}}= \{ 2(\hat\kappa_{\mathcal{F}} \hat p_1)^{1/2}
\hat \gamma_{j|\mathcal{F},1},\ldots,2(\hat\kappa_{\mathcal{F}} \hat p_H)^{1/2}$ $\hat \gamma_{j|\mathcal{F},H}\}^T$.
 Then
$T^\dr_{j|\mathcal{F}}=n( \hat\bfell^{\dr}_{j|\mathcal{F}} )^T  \hat\bfell^{\dr}_{j|\mathcal{F}}$ with
$$\hat\bfell^{\dr}_{j|\mathcal{F}}=\{ (\hat\bfell^{\dr 1}_{j|\mathcal{F}})^T, (\hat\bfell^{\dr 2}_{j|\mathcal{F}})^T, \hat\ell^{\dr 3}_{j|\mathcal{F}}, (\hat\bfell^{\dr 4}_{j|\mathcal{F}})^T,
(\hat\bfell^{\dr 5}_{j|\mathcal{F}})^T\}^T.$$
Let $ (\bfell^{\dr}_{j|\mathcal{F}})^*=(\{(\bfell^{\dr 1}_{j|\mathcal{F}})^*\}^T,\{(\bfell^{\dr 2}_{j|\mathcal{F}})^*\}^T,(\ell^{\dr 3}_{j|\mathcal{F}})^*,
  \{(\bfell^{\dr 4}_{j|\mathcal{F}})^*\}^T, \{(\bfell^{\dr 5}_{j|\mathcal{F}})^*\}^T )^T$, where
 $(\bfell^{\dr 1}_{j|\mathcal{F}})^*$ $=\{\sqrt{2}p_1^{1/2}
(1-\zeta_{j|\mathcal{F},1})^*,\ldots,\sqrt{2} p_H^{1/2}(1-\zeta_{j|\mathcal{F},H}^*)\}^T$,
 $(\bfell^{\dr 2}_{j|\mathcal{F}})^*=\{2 p_1^{1/2}(\bfnu_{j|\mathcal{F},1}^*)^T, \ldots,2 p_H^{1/2}$ $(\bfnu_{j|\mathcal{F},H}^*)^T \}^T$,
 $(\ell^{\dr 3}_{j|\mathcal{F}})^*=0$, $(\bfell^{\dr 4}_{j|\mathcal{F}})^*=2\sum_{h=1}^H  p_h$
$\bfiota_{j|\mathcal{F},h}^*$, and
 $(\bfell^{\dr 5}_{j|\mathcal{F}})^*=\{2(\kappa_{\mathcal{F}} p_1)^{1/2}(\gamma_{j|\mathcal{F},1}^*)^T,$ $\ldots, 2 (\kappa_{\mathcal{F}} p_H)^{1/2}(\gamma_{j|\mathcal{F},H}^*)^T \}^T$.
 Here  $\gamma_{j|\mathcal{F},h}^*$ is provided in Lemma \ref{lemma:one}, $\bfnu_{j|\mathcal{F},h}^*$ and $\zeta_{j|\mathcal{F},h}^*$
  are
  provided in Lemma \ref{lemma:two}, and $\bfiota_{j|\mathcal{F},h}^*$ is provided in Lemma \ref{lemma:three}.
  With $\Y \indep \x_j| \xf$,  conditions (\ref{eq:subset lcm}) and (\ref{eq:subset ccv}), we have
  $\gamma_{j|\mathcal{F},h}=0$, $\zeta_{j|\mathcal{F},h}=1$, $\bfnu_{j|\mathcal{F},h}=\0$. It follows that
$$\hat \bfell^{\dr}_{j|\mathcal{F}}=\bfell^{\dr}_{j|\mathcal{F}}+E_n\{(\bfell^{\dr}_{j|\mathcal{F}})^*\}+o_P(n^{-1/2}).$$
%where $\bfphi_{j|\mathcal{F},h}^*$ and $\zeta_{j|\mathcal{F},h}^*$
%  are
%  defined in Lemma \ref{lemma:two}.
   Define ${\bf
\Omega}^{\dr}_{j|\mathcal{F}}=E((\bfell^{\dr}_{j|\mathcal{F}})^*\{(\bfell^{\dr}_{j|\mathcal{F}})^*\}^T)$,
and the result of Theorem \ref{theorem:dr} follows directly.\eop

With the expansion forms in the proofs of Theorems \ref{theorem:sir}, \ref{theorem:save} and \ref{theorem:dr},  we construct consistent estimators for ${\bf
\Omega}^{\sir}_{j|\mathcal{F}}$, ${\bf
\Omega}^{\save}_{j|\mathcal{F}}$ and  ${\bf
\Omega}^{\dr}_{j|\mathcal{F}}$ as follows: ${\bf\hat\Omega}^\sir_{j|\mathcal{F}}=E_n((\bfell^\sir_{j|\mathcal{F}})^*\{(\bfell^\sir_{j|\mathcal{F}})^*\}^T)$,
${\bf\hat\Omega}^\save_{j|\mathcal{F}}=E_n((\bfell^\save_{j|\mathcal{F}})^*\{(\bfell^\save_{j|\mathcal{F}})^*\}^T)$, and
${\bf\hat\Omega}^\dr_{j|\mathcal{F}}=E_n((\bfell^\dr_{j|\mathcal{F}})^*\{(\bfell^\dr_{j|\mathcal{F}})^*\}^T)$.
%\begin{align}
%\label{eq:est Omega}
%\begin{split}
%{\bf\hat\Omega}^\sir_{j|\mathcal{F}}=&E_n((\bfell^\sir_{j|\mathcal{F}})^*\{(\bfell^\sir_{j|\mathcal{F}})^*\}^T),\\
%{\bf\hat\Omega}^\save_{j|\mathcal{F}}=&E_n((\bfell^\save_{j|\mathcal{F}})^*\{(\bfell^\save_{j|\mathcal{F}})^*\}^T),\\
%{\bf\hat\Omega}^\dr_{j|\mathcal{F}}=&E_n((\bfell^\dr_{j|\mathcal{F}})^*\{(\bfell^\dr_{j|\mathcal{F}})^*\}^T).
%\end{split}
%\end{align}
Then the estimated weights $\hat\omega_{j|\mathcal{F},k}^\sir$, $\hat\omega_{j|\mathcal{F},k}^\save$ and $\hat\omega_{j|\mathcal{F},k}^\dr$ are the $k$th eigenvalue of
${\bf\hat\Omega}^\sir_{j|\mathcal{F}}$, ${\bf\hat\Omega}^\save_{j|\mathcal{F}}$ and ${\bf\hat\Omega}^\dr_{j|\mathcal{F}}$ respectively.

\csubsection{Proofs of Theorems in Section 4}\label{pf sec4}

\noindent {\sc Proof of Proposition \ref{prop:max trace change}}.
For any $j\in \calf^c\cap \cala$, we know from Theorem \ref{prop:sir}
and the fact $\x_{j|\calf}=x_j-E^T(x_j\X_{\calf})\sig_{\calf}^{-1}\X_{\calf}$
 that
\begin{align}
\begin{split}
&\sigma^2_{j|\calf}\{\trace(\M^\sir_{\calf\cup j})-\trace(\M^\sir_\calf)\}=\var(\x_{j|\calf}|Y)\\
=&\{-E^T(x_j\X_{\calf})\sig_{\calf}^{-1},1\}\var\{E((\X_{\calf}^T,x_j) |Y)\}\{-E^T(x_j\X_{\calf})\sig_{\calf}^{-1},1\}^T\\
=&\{-E^T(x_j\X_{\calf})\sig_{\calf}^{-1},1\} {\bf P}\var\{E(\X|Y)\} {\bf P}^T\{-E^T(x_j\X_{\calf})\sig_{\calf}^{-1},1\}^T.
\end{split}
\label{max trace change1}
\end{align}
Here ${\bf P}=\left({\bf I}_{|\mathcal{F}|+1}, {\bf 0}_{(|\mathcal{F}|+1)\times (p-{|\mathcal{F}|-1})}\right)$, and we assume without loss of generality that the first $|\mathcal{F}|+1$ elements of $\X^T$ are $(\X_{\calf}^T,x_j)$.
Since $\M^{\sir}=\var\{E(\Z|Y)\}=\sum_{i=1}^q \lambda_i\bfeta_i\bfeta_i^T$ and $\bfbeta_i=\sig^{-1/2}\bfeta_i$, we have
\begin{align}
\var\{E(\X|Y)\}=\sig^{1/2}(\sum_{i=1}^q \lambda_i\bfeta_i\bfeta_i^T)\sig^{1/2}=\sig(\sum_{i=1}^q \lambda_i\bfbeta_i\bfbeta_i^T)\sig.
\label{max trace change2}
\end{align}
Recall that $\cali=\{1,\ldots,p\}$. Denote $\sig_{\calf_1,\calf_2}=E(\X^T_{\calf_1} \X_{\calf_2})$ for any
$\calf_1,\calf_2\subseteq \cali$. It follows
\begin{align*}
%\begin{split}
&\{-E^T(x_j\X_{\calf})\sig_{\calf}^{-1},1\} {\bf P}\sig \bfbeta_i=\{-E^T(x_j\X_{\calf})\sig_{\calf}^{-1},1\}\sig_{\calf\cup j,\cali} \bfbeta_i\\
=&\left(\sig_{j,\cali}-\sig_{j,\calf}\sig_{\calf}^{-1}\sig_{\calf,\cali}\right) \bfbeta_i
=\left( \sig_{j,\calf^c}-\sig_{j,\calf}\sig_{\calf}^{-1}\sig_{\calf,\calf^c}\right) \bfbeta_{i,\calf^c},
%\end{split}
%\label{max trace change3}
\end{align*}
where the last equality is true because $\left(\sig_{j,\calf}-\sig_{j,\calf}\sig_{\calf}^{-1}\sig_{\calf,\calf}\right)\bfbeta_{i,\calf}={\bf 0}$.
By the definition of $\cala$ in (\ref{mfvs}) and the fact that  $\spn(\bfbeta_1,\ldots,\bfbeta_q)=\spc_{\Y|\X}$,
we know $\bfbeta_{i,\calf^c\cap \cala^c}={\bf 0}$. Thus for $i=1,\ldots,q$,
\begin{align}
\{-E^T(x_j\X_{\calf})\sig_{\calf}^{-1},1\} {\bf P}\sig \bfbeta_i=
\left(\sig_{j,\calf^c\cap \cala}-\sig_{j,\calf}\sig_{\calf}^{-1}\sig_{\calf,\calf^c\cap \cala}\right)\bfbeta_{i,\calf^c\cap \cala}.
\label{max trace change3}
\end{align}
(\ref{max trace change1}), (\ref{max trace change2}) and (\ref{max trace change3}) together imply that
\begin{align*}
\sigma^2_{j|\calf}\{\trace(\M^\sir_{\calf\cup j})-\trace(\M^\sir_\calf)\}
=\sum_{i=1}^q \lambda_i\{(\sig_{j,\calf^c\cap \cala}-\sig_{j,\calf}\sig_{\calf}^{-1}\sig_{\calf,\calf^c\cap \cala})\bfbeta_{i,\calf^c\cap \cala}\}^2.
\end{align*}
By noticing that
$\sum_{j\in \calf^c\cap\cala} \{(\sig_{j,\calf^c\cap \cala}-\sig_{j,\calf}\sig_{\calf}^{-1}\sig_{\calf,\calf^c\cap \cala})\bfbeta_{i,\calf^c\cap \cala}\}^2=\bfbeta_{i,\calf^c\cap\cala}^T (\sig_{\calf^c\cap \cala}-\sig_{\calf^c\cap \cala,\calf}\sig_{\calf}^{-1}\sig_{\calf,\calf^c\cap \cala})^2\bfbeta_{i,\calf^c\cap\cala}$, and
%\begin{align*}
%& \lambda_{\min}(\sig_{\calf^c\cap \cala}-\sig_{\calf^c\cap \cala,\calf}\sig_{\calf}^{-1}\sig_{\calf,\calf^c\cap \cala})\\
%=&  \lambda_{\max}^{-1}\{(\sig_{\calf^c\cap \cala}-\sig_{\calf^c\cap \cala,\calf}\sig_{\calf}^{-1}\sig_{\calf,\calf^c\cap \cala})^{-1}\}\ge\lambda_{\max}^{-1}(\sig^{-1})=\lambda_{\min}(\sig),
%\end{align*}
$\lambda_{\min}(\sig_{\calf^c\cap \cala}-\sig_{\calf^c\cap \cala,\calf}\sig_{\calf}^{-1}\sig_{\calf,\calf^c\cap \cala})
=  \lambda_{\max}^{-1}\{(\sig_{\calf^c\cap \cala}-\sig_{\calf^c\cap \cala,\calf}\sig_{\calf}^{-1}\sig_{\calf,\calf^c\cap \cala})^{-1}\}\ge\lambda_{\max}^{-1}(\sig^{-1})=\lambda_{\min}(\sig)$,
we have
\begin{align*}
&\max_{j\in \calf^c\cap\cala} \sigma^2_{j|\calf}\{\trace(\M^\sir_{\calf\cup j})-\trace(\M^\sir_\calf)\}
\ge |\calf^c\cap\cala|^{-1} \sum_{j\in \calf^c\cap\cala} \left[\sigma^2_{j|\calf}\{\trace(\M^\sir_{\calf\cup j})-\trace(\M^\sir_\calf)\} \right]\\
 &\hspace{.2in}= |\calf^c\cap\cala|^{-1} \sum_{i=1}^q \lambda_i \bfbeta_{i,\calf^c\cap\cala}^T (\sig_{\calf^c\cap \cala}-\sig_{\calf^c\cap \cala,\calf}\sig_{\calf}^{-1}\sig_{\calf,\calf^c\cap \cala})^2\bfbeta_{i,\calf^c\cap\cala}\\
 &\hspace{.2in}\ge |\calf^c\cap\cala|^{-1} \sum_{i=1}^q \lambda_i \lambda^2_{\min}(\sig_{\calf^c\cap \cala}-\sig_{\calf^c\cap \cala,\calf}\sig_{\calf}^{-1}\sig_{\calf,\calf^c\cap \cala})\bfbeta_{i,\calf^c\cap\cala}^T\bfbeta_{i,\calf^c\cap\cala}\\
 &\hspace{.2in}\ge  \lambda_q \lambda^2_{\min}(\sig) |\calf^c\cap\cala|^{-1}  \sum_{i=1}^q \bfbeta_{i,\calf^c\cap\cala}^T\bfbeta_{i,\calf^c\cap\cala}\ge
 \lambda_q \lambda^2_{\min}(\sig) \bfbeta^2_{\min}.
%\ge d \lambda_d \lambda^{-3}_{\max}(\sig)\bfbeta^2_{\min}
\end{align*}
The proof is then completed by noting that
$\max_{j\in \calf^c\cap\cala}\{\trace(\M^\sir_{\calf\cup j})-\trace(\M^\sir_\calf)\}\ge
\max_{j\in \calf^c\cap\cala}$ $\sigma^2_{j|\calf}\{\trace(\M^\sir_{\calf\cup j})-\trace(\M^\sir_\calf)\}/\max_{j\in \calf^c\cap\cala}\sigma^2_{j|\calf}$
and  $\sigma_{j|\calf}^2\le \var(x_j)\le \lambda_{\max}(\sig)$.
\eop

 \noindent {\sc Proof of Theorem \ref{theorem:selection}.}
%   We first discuss condition (\ref{eq:sep}). From Theorem \ref{prop:sir}, $\trace(\M^\sir_{\calf\cup j})-\trace(\M^\sir_\calf)=\sum_{h=1}^H p_h
%\gamma_{j|\mathcal{F},h}^2$. Note that $\sum_{h=1}^H p_h=1$, together with condition (\ref{eq:sep}), we have
%\begin{align}
%\label{eq:sep sir}
%\underset
%{\substack{
%            {\tiny \calf :  \calf^c\cap \cala\neq \varnothing}\\
%             {\tiny j\in\calf^c} }}
%{\mbox{min }}
%\trace\big(\M^\sir_{\calf\cup j}\big)
%-\trace\big( \M^\sir_{{\mathcal{F}}}\big) >\sum_{h=1}^H p_h \varsigma n^{-\xi_{\min}}=\varsigma n^{-\xi_{\min}}.
%\end{align}
%We could replace condition   (\ref{eq:sep}) in Theorem  \ref{theorem:selection} with  the weaker condition (\ref{eq:sep sir}).
 For part 1, denote
 $\Delta=\varsigma n^{-\xi_{\min}}-n^{-1}\bar c^\sir>0$. Because $0<\bar c^\sir<\varsigma n^{1-\xi_{\min}}/2$, we have
 $\Delta=O_P(n^{-\xi_{\min}})$. When $\calf^c\cap\cala\neq \varnothing$,
 $\left\{\trace\big(\hat \M^\sir_{\calf\cup j}\big)
-\trace\big(\hat \M^\sir_{{\mathcal{F}}}\big)\right\}-\left\{\trace\big( \M^\sir_{\calf\cup j}\big)
-\trace\big( \M^\sir_{{\mathcal{F}}}\big)\right\}=O_P(n^{-1/2})$.
 % $\bar c^\sir<\varsigma n^{1-\xi_{\min}}$ implies that
%
%  As $n$ goes to infinity, $\trace\big(\hat \M^\sir_{\calf\cup j}\big)
%-\trace\big(\hat \M^\sir_{{\mathcal{F}}}\big)$ converges to $\trace\big( \M^\sir_{\calf\cup j}\big)
%-\trace\big( \M^\sir_{{\mathcal{F}}}\big)$.
Note that $0<\xi_{\min}<1/2$. Thus as $n$ goes to infinity, with probability approaching 1,
 $$\max_{\calf:\calf^c\cap\cala\neq \varnothing} \max_{j \in \calf^c\cap \cala} \left[ \left\{\trace\big( \M^\sir_{\calf\cup a_\calf}\big)
-\trace\big( \M^\sir_{{\mathcal{F}}}\big) \right\}-\left\{\trace\big( \hat\M^\sir_{\calf\cup a_\calf}\big)
-\trace\big( \hat\M^\sir_{{\mathcal{F}}}\big)\right\}\right]<\Delta.$$
  Together with (\ref{eq:sep}), we know with probability approaching 1,
  \begin{align*}
  &\min_{\calf:\calf^c\cap\cala\neq \varnothing} \max_{j \in \calf^c\cap \cala}
  \left\{\trace\big( \hat\M^\sir_{\calf\cup a_\calf}\big)
-\trace\big( \hat\M^\sir_{{\mathcal{F}}}\big)\right\}
%\\
%&\hspace{.2in}
>  \min_{\calf:\calf^c\cap\cala\neq \varnothing} \max_{j \in \calf^c\cap \cala}
  \left\{\trace\big( \M^\sir_{\calf\cup a_\calf}\big)
-\trace\big( \M^\sir_{{\mathcal{F}}}\big)\right\}\\
&\hspace{.2in}- \max_{\calf:\calf^c\cap\cala\neq \varnothing} \max_{j \in \calf^c\cap \cala} \left[ \left\{\trace\big( \M^\sir_{\calf\cup a_\calf}\big)
-\trace\big( \M^\sir_{{\mathcal{F}}}\big) \right\}-\left\{\trace\big( \hat\M^\sir_{\calf\cup a_\calf}\big)
-\trace\big( \hat\M^\sir_{{\mathcal{F}}}\big)\right\}\right]\\
&\hspace{.2in}>  \varsigma n^{-\xi_{\min}}-  \Delta=n^{-1}\bar c^\sir.
  \end{align*}
Multiply both sides by $n$ and
 we get $Pr(\min_{\calf:\calf^c\cap\cala\neq \varnothing} \max_{j \in \calf^c\cap \cala} T^{\textup{SIR}}_{j|\mathcal{F}}>
   \bar c^\sir )\rightarrow 1$.
% $Pr(\underset{\calf :  \calf^c\cap \cala\neq \varnothing}{\mbox{min }}T^{\textup{SIR}}_{a_\calf|\mathcal{F}}>
%   \bar\delta_{a_\calf|\mathcal{F}}^\sir )\rightarrow 1$.

 In part 2, note that $\calf^c\cap\cala= \varnothing$ implies $\cala\subseteq\calf$. Then
  $j\in \calf$ implies either $j\in \cala$ or $j\in \{\calf\backslash \cala\}$. If $j\in \cala$,  then $\{{\mathcal{F}}\backslash {j}\}^c\cap \cala\neq \varnothing$.
  As $n$ goes to infinity, $\trace(\hat \M^\sir_\calf\big)
-\trace\big(\hat \M^\sir_{\{{\mathcal{F}}\backslash{j}\}}\big)$ converges to $\trace\big(\M^\sir_\calf\big)
-\trace\big(\M^\sir_{\{{\mathcal{F}}\backslash{j}\}}\big)$.
    Condition (\ref{eq:sep}) implies that  for any $j\in \cala$,
 $T^{\textup{SIR}}_{j|\{{\mathcal{F}}\backslash{j}\}}>\varsigma n^{1-\xi_{\min}}/2$ with probability 1.
If $j\in \{\calf\backslash \cala\}$,
 Theorem \ref{theorem:sir} guarantees that $T^{\textup{SIR}}_{j|\{{\mathcal{F}}\backslash{j}\}}$ converges to a sum of weighted $\chi^2$, which is $O_p(1)$ and is asymptotically smaller than $\varsigma n^{1-\xi_{\min}}/2$.
 Thus for $\calf^c\cap\cala= \varnothing$, $\min_{j \in \calf} T^{\textup{SIR}}_{j|\{{\mathcal{F}}\backslash{j}\}}=O_P(1)<C n^{1-\xi_{\min}}$ with $\xi_{\min}<1$ and $C>0$. It follows that $Pr(\max_{\calf:\calf^c\cap\cala= \varnothing} \min_{j \in \calf} T^{\textup{SIR}}_{j|\{{\mathcal{F}}\backslash j\}}<
\underline{c}^{\textup{SIR}} )\rightarrow 1$ if we set $\underline{c}^{\textup{SIR}}>C n^{1-\xi_{\min}}$.
 \eop

For stepwise SAVE, we replace condition (\ref{eq:sep}) with
\begin{align}
\label{eq:save sep}
\min_{\calf:\calf^c\cap\cala\neq \varnothing} \max_{j \in \calf^c\cap \cala} \{ \trace(\M^\save_{\calf\cup j})-\trace(\M^\save_\calf)\}>\varsigma n^{-\xi_{\min}}.
%\mbox{ for some } \varsigma>0 \mbox{ and } 0\le\xi_{\min}<1.
\end{align}
For stepwise directional regression, we replace condition (\ref{eq:sep}) with
\begin{align}
\label{eq:dr sep}
\min_{\calf:\calf^c\cap\cala\neq \varnothing} \max_{j \in \calf^c\cap \cala} \{ \trace(\M^\dr_{\calf\cup j})-\trace(\M^\dr_\calf)\}>\varsigma n^{-\xi_{\min}}.
%\mbox{ for some } \varsigma>0 \mbox{ and } 0\le\xi_{\min}<1.
\end{align}
Conditions (\ref{eq:save sep}) and (\ref{eq:dr sep}) are parallel to condition (\ref{eq:sep}), and will guarantee the selection consistency of stepwise SAVE and stepwise directional regression respectively.

Before proceeding to the proof of Theorem \ref{theo:sure screening}, we present the following useful lemmas.
For ${\bf a}=(a_1,\ldots,a_p)^T\in\R^p$, let $\|{\bf a}\|_\infty=\max_{1\leq i\leq p}|a_i|$,
$\|{\bf a}\|_1=\sum_{i=1}^p |a_i|$,
and $\|{\bf a}\|_2=\sqrt{\sum_{i=1}^p a_i^2}$. For ${\bf A}=\{a_{ij}\}\in\R^{p\times p}$, let $\|{\bf A}\|_{\infty}=\max_{1\le i,j\le p} |a_{ij}|$.

\begin{lemma}\label{lemma: ss prep}
For $h=1,\ldots,H$, let ${W}_h=p_h^{-1/2}\left(I(\Y\in J_h)-p_h\right)$ and
${\ba}_h=\sig^{-1}E(\X{W}_h)$. Then $\trace(\M^{\sir})=(H-1)-\sum_{h=1}^H E ({W}_h-\X^T \ba_h)^2$.
\end{lemma}
\noindent {\sc Proof of Lemma \ref{lemma: ss prep}}.
%The first part is obvious. For the second part,
 We notice that for $h=1,\ldots,H$,
\begin{align*}
&E({W}_h- \X^T {\ba_h})^2 =  E ({W}_h^2)+ E({\ba_h}^T \X \X ^T {\ba_h})- 2 E ({W}_h \X^T \ba_h)\\
&= E ({W}_h^2) - {\ba_h}^T \sig {\ba_h}
= (1-p_h)- p_h^{-1} E\{\Z^T I(\Y \in J_h)\}E\{\Z I(\Y \in J_h)\}.
\end{align*}
Add over $h$ and we get the desired result.
\eop

\begin{lemma}\label{lemma:four}
Let $\W^{\sir}=\var\{ E(\X|\Y)\}$  and let $\hat{\W}^{\sir}$ be its corresponding sample estimator.
Denote $p_0=\min_{1\leq h \leq H} p_h$, $\tau_0=1.25\exp\{1+(10\tau_{\min})^{-1/2}\tau_{\max}^{1/2}\}$, $D_1=2+8\tau_0^2$, $D_2=(10\tau_{\min})^{1/2}D_1$, $D_3=2H\tau_{\max}^{1/2} p_0^{-1}D_2$ and $D_4=D_3+Hp_0^{-1}D_2^2+(2+12.5e^2)^2$. Then under the same conditions of Theorem \ref{theo:sure screening}, we have
%\begin{align*}
$\max_{1\le i,j\le p} |\hat{\W}^\sir_{ij}-{\W}^\sir_{ij}|\le D_4 \sqrt{\log p /n }$
%\end{align*}
with probability tending to $1$.
\end{lemma}

\noindent {\sc Proof of Lemma \ref{lemma:four}.} Let $\calu_h= E\{\X I(\Y\in J_h)\}$ and let $\hat{\calu}_h=E_n\{\X I(\Y\in J_h)\}$ be its sample estimator. By the definition of $\W^{\sir}$ and $\hat{\W}^{\sir}$, we have
\begin{align*}
\hat{\W}^{\sir}-\W^{\sir}=\sum_{h=1}^H \{\hat p_h^{-1} \hat\calu_h\hat{\calu}_h - p_h^{-1}\calu_h\calu_h^T \}- E_n(\X)E_n(\X^T).
\end{align*}
Denote $\W^{(1)}=\sum_{h=1}^H  p_h^{-1} (\hat\calu_h-\calu_h){\calu}_h^T$,  $\W^{(2)}=\sum_{h=1}^H  p_h^{-1} \calu_h(\hat\calu_h-\calu_h)^T$, $\W^{(3)}=
\sum_{h=1}^H  p_h^{-1} (\hat\calu_h-\calu_h)(\hat\calu_h-\calu_h)^T$, $\W^{(4)}=\sum_{h=1}^H (\hat p_h p_h)^{-1}(\hat p_h-p_h)\hat{\calu}_h\hat{\calu}_h^T$, and $\W^{(5)}=E_n(\X)E_n(\X^T)$. Then we have
\begin{align}
\hat{\W}^{\sir}-\W^{\sir}=
\W^{(1)}+\W^{(2)}+\W^{(3)}+\W^{(4)}-\W^{(5)}.
\label{eq: expan w}
\end{align}

Since $\X=(x_1,\ldots,x_p)^T$ is normal, condition (C2) implies that $E\{\exp(tx_i^2)\}\le 1.25$ for any $t$ such that $0\le t\le (10\tau_{\min})^{-1}$.
 Inequality $\exp(s)\le \exp(s^2+1)$ implies that $E\{\exp(t|x_i|)\}\le 1.25e$ as long as $0\le t\le (10\tau_{\min})^{-1}$.
   Note that
$|x_iI(\Y\in J_h)|\le |x_i|$ and $|\calu_{h,i}|=|E\{x_i I(Y\in J_h)\}|\le E(x_i^2)^{1/2}\le \tau_{\max}^{1/2}$. We have
%\begin{align*}
$E\{\exp(t|x_i I(\Y\in J_h)-\calu_{h,i}|)\}\le  E{\exp(t|x_i|)}\exp\{(10\tau_{\min})^{-1/2}\tau_{\max}^{1/2}\}
\le  1.25\exp\{1+(10\tau_{\min})^{-1/2}\tau_{\max}^{1/2}\}$
%\end{align*}
for $0\le t\le (10\tau_{\min})^{-1/2}$.
 Let $\epsilon=(10\tau_{\min})^{-1/2}\sqrt{\log p/n}$.
 Following similar arguments in the proof of Theorem 1 in Cai et al. (2011), we have
\begin{align*}
\qquad\quad &Pr(\|\hat{\calu}_{h}-\calu_{h}\|_{\infty}\ge D_2\sqrt{\log p/n})\\
\le& 2p\exp(-D_1\log p) [E\exp\{\epsilon(x_i I(\Y\in J_h)-\calu_{h,i})\}]^n \\
\le &2p\exp\{-D_1\log p + n\epsilon^2E(x_i I(\Y\in J_h)-\calu_{h,i})^2\exp(\epsilon|x_i I(\Y\in J_h)-\calu_{h,i}|)\}\\
\le &2p\exp\{-D_1\log p + 8\tau_0^2\log p\}=2p\exp\{-2\log p\}.
%\\\label{eq: bound 1}
%=& 2p^{-1}.
\end{align*}
Thus we have
\begin{align}
Pr(\|\hat{\calu}_{h}-\calu_{h}\|_{\infty}\ge D_2\sqrt{\log p/n})\le 2p^{-1}.
\label{eq: bound 1}
\end{align}
%For $i=1,\ldots,q$,  we have $\tau_{\min}\|\bfbeta_i\|^2_{2}\le\bfbeta_i^T\sig\bfbeta_i=1$ and
%$\|\bfbeta_i\|^2_{1}\le |
%\cala|\|\bfbeta_i\|^2_{2} \le \varsigma n^{\xi_0} \tau_{\min}^{-1}$.
%With condition (\ref{eq:lcm}), we have
% \begin{align}\nonumber
% \|\U_h\|_{1}=&\|\sig (\bfbeta_1,\ldots,\bfbeta_q)(\bfbeta_1,\ldots,\bfbeta_q)^T\U_h\|_{1}\\\nonumber
% \le & q \tau_{\max} \max_{1\le i\le q} \| \bfbeta_i\|^2_{1} \|\U_h\|_{\infty}
% \le q \tau_{\max} (\varsigma n^{\xi_0} \tau_{\min}^{-1}) (\tau_{\max}^{1/2} p_h^{-1})\\\label{eq: bound 2}
% \le & q \varsigma p_0^{-1} \tau_{\min}^{-1}\tau_{\max}^{3/2} n^{\xi_0}.
% \end{align}
Moreover, it is easy to see that
\begin{align}\label{eq: bound 2}
\|\U_h\|_{\infty}\le \tau_{\max}^{1/2} p_h^{-1}\le \tau_{\max}^{1/2} p_0^{-1}.
\end{align}
  Combining (\ref{eq: bound 1}) and (\ref{eq: bound 2}), we see with probability tending to $1$,
\begin{align}\label{eq: bound 3}
\|\W^{(1)}+\W^{(2)}\|_{\infty}\le D_3 \sqrt{\log p/n}.
\end{align}
By (\ref{eq: bound 1}) and condition (C4), we see with probability tending to $1$,
\begin{align}\label{eq: bound 4}
\|\W^{(3)}\|_{\infty}\le H p_0^{-1} D_2^2 \sqrt{\log p/n}.
\end{align}
Similar to (\ref{eq: bound 1}), we can also show that
%\begin{align*}
$Pr(\|E_n(\X)\|_{\infty} \ge \{2+ 8(1.25e)^2\}\sqrt{\log p/n})\le 2p^{-1}$.
%\end{align*}
Under condition (C3),  we have with probability tending to $1$,
 \begin{align}\label{eq: bound 5}
 \|\W^{(5)}\|_{\infty} \le (2+ 12.5e^2)^2\sqrt{\log p/n}.
 \end{align}
Because $\hat p_h- p_h= O_P(n^{-1/2})$, we know that $\|\W^{(4)}\|_{\infty}=O_P(n^{-1/2})$.
 Together with (\ref{eq: expan w}), (\ref{eq: bound 3}), (\ref{eq: bound 4}), and (\ref{eq: bound 5}), we get the desired result.
 \eop

 %{\bf The following Lemma is given in Cai et al. (2011) and thus detail proof is omitted.}
\begin{lemma}\label{lemma:five}
Assume conditions \textup{(C1)} and \textup{(C2)} hold. Then there exists $D_5>0$ such that $\max_{1\le i,j\le p} |\hat\sig_{ij}-\sig_{ij}|\le D_5 \sqrt{\log p/n}$ with probability tending to 1.
\end{lemma}
\noindent The proof of this Lemma is available in Cai et al. (2011) and thus omitted.

\begin{lemma}\label{lemma:six}
Let $\bfdelta_{j|\calf}=(-\bftheta_{j|\calf}^T,1)^T$ and $\hat{\bfdelta}_{j|\calf}=(-\hat{\bftheta}_{j|\calf}^T,1)^T$. Define $D_6=1+ 16 \tau_{\min}^{-2} (\tau_{\max})^2$ and $D_0=D_4D_6+(4\tau_{\min}^{-2}\tau^2_{\max}+\tau_{\min}^{-1}\tau_{\max}) D_5D_6$. Assume the same conditions of Theorem \ref{theo:sure screening} hold. Suppose $|\calf|=O(n^{\xi_0+\xi_{\min}})$.
Then with probability tending to 1, we have
%\begin{align}\nonumber
$|\hat{\bfdelta}^T_{j|\calf}\hat\W^{\sir}_{\calf\cup j}\hat{\bfdelta}_{j|\calf} - {\bfdelta}^T_{j|\calf}\W^{\sir}_{\calf\cup j}{\bfdelta}_{j|\calf}|
\le D_0|\calf|\sqrt{\log p/n}$.
%\end{align}
\end{lemma}

\noindent {\sc Proof of Lemma \ref{lemma:six}.} For any $|\calf|=O(n^{\xi_0+\xi_{\min}})$, we know from Lemm 1 in Wang (2009) that
$2^{-1}\tau_{\min}<\lambda_{\min}(\hat{\sig}_{\calf})<\lambda_{\max}(\hat{\sig}_{\calf})<2\tau_{\max}$.
Moreover, $\lambda_{\max}({\W}^{\sir}_{\calf})\le \lambda_{\max}({\sig}_{\calf})<\tau_{\max}$ and $\lambda_{\max}(\hat{\W}^{\sir}_{\calf})\le \lambda_{\max}(\hat{\sig}_{\calf})<2\tau_{\max}$.
It follows that
\begin{align}
\label{eq: bound 6}
\begin{split}
\|{\bfdelta}_{j|\calf}\|^2_{2}&=1+E(x_j\X^T_{\calf})\sig^{-2}_{\calf}E(x_j\X_{\calf})\le 1+ \tau_{\min}^{-2}\|E(x_j\X_{\calf})\|^2_{2}\\
&\le 1+ \tau_{\min}^{-2} (\tau_{\max})^2,
\end{split}
\end{align}
and
\begin{align}
\label{eq: bound 7}
\begin{split}
\|\hat{\bfdelta}_{j|\calf}\|^2_{2}&=1+E_n(x_j\X^T_{\calf})\hat\sig^{-2}_{\calf}E_n(x_j\X_{\calf})\le 1+ 4\tau_{\min}^{-2}\|E_n(x_j\X_{\calf})\|^2_{2}\\
&\le 1+ 16 \tau_{\min}^{-2} (\tau_{\max})^2.
\end{split}
\end{align}
By triangular inequality, we have
\begin{align}\label{eq: bound 8}
\begin{split}
&|\hat{\bfdelta}^T_{j|\calf}\hat\W^{\sir}_{\calf\cup j}\hat{\bfdelta}_{j|\calf} - {\bfdelta}^T_{j|\calf}\W^{\sir}_{\calf\cup j}{\bfdelta}_{j|\calf}| \le |\hat{\bfdelta}^T_{j|\calf}\hat\W^{\sir}_{\calf\cup j}\hat{\bfdelta}_{j|\calf} - \hat{\bfdelta}^T_{j|\calf}\W^{\sir}_{\calf\cup j}\hat {\bfdelta}_{j|\calf}| \\
&
+|\hat{\bfdelta}^T_{j|\calf}\W^{\sir}_{\calf\cup j}\hat{\bfdelta}_{j|\calf} - {\bfdelta}^T_{j|\calf}\W^{\sir}_{\calf\cup j}{\bfdelta}_{j|\calf}|
\end{split}
\end{align}
We bound the two terms of (\ref{eq: bound 8}) respectively. Invoking Lemma \ref{lemma:four} and (\ref{eq: bound 7}), we have
\begin{align}
\label{eq: bound 9}
\begin{split}
& |\hat{\bfdelta}^T_{j|\calf}\hat\W^{\sir}_{\calf\cup j}\hat{\bfdelta}_{j|\calf} - \hat{\bfdelta}^T_{j|\calf}\W^{\sir}_{\calf\cup j}\hat {\bfdelta}_{j|\calf}|
\le  \|\hat{\bfdelta}^T_{j|\calf}\|^2_{\ell_2} |\calf| \max_{1\le i,j\le p} |\hat{\W}^\sir_{ij}-{\W}^\sir_{ij}|\\
&\le  D_4D_6|\calf| \sqrt{\log p/n},
\end{split}
\end{align}
and
\begin{align}
\label{eq: bound 13}
\begin{split}
&|\hat{\bfdelta}^T_{j|\calf}\W^{\sir}_{\calf\cup j}\hat{\bfdelta}_{j|\calf} - {\bfdelta}^T_{j|\calf}\W^{\sir}_{\calf\cup j}{\bfdelta}_{j|\calf}|
= |(\hat{\bfdelta}^T_{j|\calf}+{\bfdelta}^T_{j|\calf})^T\W^{\sir}_{\calf\cup j}(\hat{\bfdelta}^T_{j|\calf}-{\bfdelta}^T_{j|\calf})|\\
&\le  \tau_{\max} \|\hat{\bfdelta}^T_{j|\calf}+{\bfdelta}^T_{j|\calf}\|_{2} \|\hat{\bfdelta}^T_{j|\calf}-{\bfdelta}^T_{j|\calf}\|_{2}.
\end{split}
\end{align}
By (\ref{eq: bound 6}) and(\ref{eq: bound 7}), we have
\begin{align}\label{eq: bound 10}
\|\hat{\bfdelta}^T_{j|\calf}+{\bfdelta}^T_{j|\calf}\|_{2}
\le (\|\hat{\bfdelta}^T_{j|\calf}\|_{2}+\|{\bfdelta}^T_{j|\calf}\|_{2})
\le D_6.
\end{align}
Invoking Lemma \ref{lemma:four} and \ref{lemma:five}, we can derive that
\begin{align}\nonumber
&\|\hat{\bfdelta}^T_{j|\calf}-{\bfdelta}^T_{j|\calf}\|_{2}
=  \|\hat\sig^{-1}_{\calf}E_n(x_j\X_{\calf})-\sig^{-1}_{\calf}E(x_j\X_{\calf})\|_{2}\\\nonumber
\le & \|\hat\sig^{-1}_{\calf}E_n(x_j\X_{\calf})-\sig^{-1}_{\calf}E_n(x_j\X_{\calf})\|_{2}
+\|\sig^{-1}_{\calf}E_n(x_j\X_{\calf})-\sig^{-1}_{\calf}E(x_j\X_{\calf})\|_{2}\\\nonumber
\le & \lambda_{\max}^{1/2}\{\big(\hat\sig^{-1}_{\calf}-\sig^{-1}_{\calf}\big)^2\}\|E_n(x_j\X_{\calf})\|_{2}
+\lambda_{\max}\big(\sig^{-1}_{\calf}\big)|\calf|^{1/2}\|E_n(x_j\X_{\calf})-E(x_j\X_{\calf})\|_{\infty}\\\nonumber
\le & 2\tau_{\min}^{-2}|\calf|D_5\sqrt{\log p/n}\cdot2\tau_{\max}+ \tau_{\min}^{-1}|\calf|^{1/2}D_5\sqrt{\log p/n}.
\end{align}
It follows that
\begin{align}\label{eq: bound 11}
\|\hat{\bfdelta}^T_{j|\calf}-{\bfdelta}^T_{j|\calf}\|_{2}\le & (4\tau_{\min}^{-2}\tau_{\max}+\tau_{\min}^{-1}) D_5\sqrt{\log p/n}.
\end{align}
(\ref{eq: bound 10}) and (\ref{eq: bound 11}) together suggest that
%\begin{align}\nonumber
%&|\hat{\bfdelta}^T_{j|\calf}\W^{\sir}_{\calf\cup j}\hat{\bfdelta}_{j|\calf} - {\bfdelta}^T_{j|\calf}\W^{\sir}_{\calf\cup j}{\bfdelta}_{j|\calf}|\\\label{eq: bound 12}
%\le & (4\tau_{\min}^{-2}\tau_{\max}^2+\tau_{\min}^{-1}\tau_{\max}) D_5D_6|\calf|\sqrt{\log p/n}..
%\end{align}
\begin{align}
\label{eq: bound 12}
|\hat{\bfdelta}^T_{j|\calf}\W^{\sir}_{\calf\cup j}\hat{\bfdelta}_{j|\calf} - {\bfdelta}^T_{j|\calf}\W^{\sir}_{\calf\cup j}{\bfdelta}_{j|\calf}|
\le  (4\tau_{\min}^{-2}\tau_{\max}^2+\tau_{\min}^{-1}\tau_{\max}) D_5D_6|\calf|\sqrt{\log p/n}.
\end{align}
Plug  (\ref{eq: bound 9}), (\ref{eq: bound 13}) and (\ref{eq: bound 12}) into (\ref{eq: bound 8}), and we get the desired result.
\eop

\noindent {\sc Proof of Theorem \ref{theo:sure screening}.} %Our proof relies heavily on the proof of Theorem 1 in Wang (2009), with two major differences. First, Wang (2009)'s result is based on forward regression in linear models. From Proposition \ref{prop: ss prep}, SIR based forward trace pursuit (FTP) can be viewed as sum of $H$ linear models. Second, the responses $W_h$ in SIR based FTP involves $p_h$, which has to be estimated at the sample level, while the response in Wang (2009)'s forward regression is observable at the sample level.
Let $C_0=2H\varsigma^{-1}$. We first state the outline of the proof as follows. Notice that $|\cala|\le \varpi n^{\xi_0}$ from condition (C4). To include $|\cala|$ relevant predictors in the FTP algorithm within
$[C_0 \varpi n^{\xi_0+\xi_{\min}}]$ steps, all we need to show is that within  $[C_0 n^{\xi_{\min}}]$ steps, at least one new significant
 variable will be selected by the FTP algorithm, conditional on those already included.
 A complete proof would entail $|\cala|$ stages, with each stage focusing on the $i$th block of $[C_0 n^{\xi_{\min}}]$ steps in the FTP algorithm, $i=1,\ldots,|\cala|$.
  Without loss of generality,
 we focus on the first block of $[C_0 n^{\xi_{\min}}]$ steps in the FTP algorithm, and show that at least one significant
 variable will be included.

 Assume no relevant predictors have been selected in the first $k$ steps,
 and we evaluate what happens at the $k+1$th step. Define
  \begin{align*}
  %\label{ftp 1}
\Omega(k)=\tr\left(\hat\M^{\sir}_{\mathcal{S}^{(k)}}\right)- \tr\left(\hat\M^{\sir}_{\mathcal{S}^{(k-1)}}\right),
k=1,2,\ldots,[C_0 n^{\xi_{\min}}].
  \end{align*}
  From this definition, we have
  %\begin{align*}
  $\sum_{k=1}^{[C_0 n^{\xi_{\min}}]}\Omega(k)=
  \tr\left(\hat\M^{\sir}_{\mathcal{S}^{([C_0 n^{\xi_{\min}}]+1)}}\right)- \tr\left(\hat\M^{\sir}_{\mathcal{S}^{(0)}}\right)$.
  %\end{align*}
  Because $\mathcal{S}^{(0)}=\varnothing$, it follows from Lemma \ref{lemma: ss prep} that
    \begin{align}\label{ftp 1}
  \sum_{k=1}^{[C_0 n^{\xi_{\min}}]}\Omega(k)=
  \tr\left(\hat\M^{\sir}_{\mathcal{S}^{([C_0 n^{\xi_{\min}}])}}\right)\leq H-1.
  \end{align}
  We will see later that
  \begin{align}\label{ftp 2}
\Omega(k)\geq  \varsigma n^{-\xi_{\min}}/2, \mbox{ if }a_{k}\notin \cala, \quad k=1,2,\ldots,[C_0 n^{\xi_{\min}}],
\end{align}
which implies
    \begin{align}\label{ftp 3}
  \sum_{k=1}^{[C_0 n^{\xi_{\min}}]}\Omega(k)\geq H \mbox{ if }a_{k}\notin \cala, \quad k=1,2,\ldots,[C_0 n^{\xi_{\min}}].
  \end{align}
  Together, (\ref{ftp 1}) and (\ref{ftp 3}) imply that there must exist $a_k\in \cala$ for some $k$ such that $1\le k \le [C_0 n^{\xi_{\min}}]$.

It remains to prove (\ref{ftp 2}).  By Theorem \ref{theorem:sir}, we can derive that
%\begin{align*}
$\trace(\M^\sir_{\calf\cup j})-\trace(\M^\sir_\calf)=\sum_{h=1}^H p_h
\gamma_{j|\mathcal{F},h}^2=\sigma^{-2}_{j|\calf}{\bfdelta}^T_{j|\calf}\W^{\sir}_{\calf\cup j}{\bfdelta}_{j|\calf}$
%\end{align*}
for any $\calf$ such that $|\calf|<n$.
In the sample level, we then have
%\begin{align*}
$\trace(\hat\M^\sir_{\calf\cup j})-\trace(\hat\M^\sir_\calf)=\hat\sigma^{-2}_{j|\calf}\hat{\bfdelta}^T_{j|\calf}\hat\W^{\sir}_{\calf\cup j}\hat{\bfdelta}_{j|\calf}$.
%\end{align*}
From the proof of Lemma 3 in Jiang and Liu (2013) and Lemma \ref{lemma:five}, we know that $\hat\sigma^{2}_{j|\calf}- \sigma^{2}_{j|\calf}=O(|\calf|\sqrt{\log p/n})$. Then under condition (C3), we see that $\hat\sigma^{-2}_{j|\calf}\ge \sigma^{-2}_{j|\calf}/2$ provided that $|\calf|=O(n^{\xi_0+\xi_{\min}})$.
Note that $|\mathcal{S}^{(k)}|\le C_0\varpi n^{\xi_0+\xi_{\min}}$. Then by Lemma \ref{lemma:six} and condition (C3), we can get
\begin{align}\nonumber
\Omega(k) \ge & 2^{-1}\sigma^{-2}_{a_{k+1}|\mathcal{S}^{(k)}} ({\bfdelta}^T_{a_{k+1}|\mathcal{S}^{(k)}}\W^{\sir}_{\mathcal{S}^{(k+1)}}{\bfdelta}_{a_{k+1}|\mathcal{S}^{(k)}}
-D_0|\mathcal{S}^{(k+1)}|\sqrt{\log p/n})\\\nonumber
\ge &(\varsigma n^{-\xi_{\min}}/2- 2^{-1}\sigma^{-2}_{a_{k+1}}D_0 \cdot (C_0\varpi) n^{(\xi_0+\xi_{\min})} \cdot\varpi^{1/2} n^{\xi/2} n^{-1/2})
\rightarrow  \varsigma n^{-\xi_{\min}}/2,
\end{align}
if $a_{k}\notin \cala$, $k=1,2,\ldots,[C_0 n^{\xi_{\min}}]$. The proof is completed. \eop

\noindent {\sc Proof of Theorem \ref{theo:BIC}.} Define $k_{\min}= \min_{1\le k \le n}\{k: \cala \subset \cals^{(k)}\}$. Theorem \ref{theo:sure screening} guarantees that $k_{\min} \le 2H\varsigma^{-1}\varpi n^{\xi_0+\xi_{\min}}$. Following the proof of Theorem 2 in Wang (2009), it's easy to prove that
%\begin{align*}
$Pr\Big(\min_{0\le k < k_{\min}} \{\textup{BIC}(\cals^{(k})- \textup{BIC}(\cals^{(k+1)})\}  >0 \Big) \rightarrow 1$,
%\end{align*}
and the details are omitted. \eop

\end{document}